\documentclass[longauth]{aa}
\pdfoutput=1
\usepackage{mathtools}
\usepackage{multirow}
\usepackage{comment}
\usepackage{siunitx}
\usepackage{booktabs}
\usepackage{graphicx}
\usepackage{menukeys}
\usepackage{algorithm}
\usepackage{ulem}
\usepackage[noend]{algpseudocode}
\usepackage{ccaption}
\usepackage{caption}
\usepackage{txfonts}
\usepackage{subfigure}
\usepackage{cuted}
\usepackage{float}
\usepackage{multicol}
\usepackage{placeins}
\usepackage{lscape}
\usepackage{soul}
\setstcolor{red}

\graphicspath{{figs/}}
\usepackage[colorlinks=true, allcolors=blue]{hyperref}
\DeclareCaptionFormat{continued}{ #1~(continued)}
\usepackage[]{starfont}
\usepackage{amsmath}

\DeclareSymbolFont{starfontsym}{OT1}{sts}{m}{n}
\DeclareMathSymbol{\mathTerra}{\mathord}{starfontsym}{76}

\DeclareSIUnit \lsun {L_\odot}
\DeclareSIUnit \rsun {R_\odot}
\DeclareSIUnit \msun {M_\odot}
\DeclareSIUnit \mearth {M_\oplus}
\DeclareSIUnit \rearth {R_\oplus}
\newcommand{\HD}{HD\,86226\,c\xspace}
\DeclareSIUnit \year {yr}
\DeclareSIUnit \day {day}
\DeclareSIUnit \pixel {px}
\DeclareSIUnit \ppm {ppm}

\newcommand{\U}{\mathrm}

\makeatletter
\renewcommand*\aa@pageof{, page \thepage{} of \pageref*{LastPage}}
\makeatother
\newcolumntype{L}[1]{>{\raggedright\arraybackslash}p{#1}}
\newcolumntype{C}[1]{>{\centering\arraybackslash}p{#1}}
\newcolumntype{R}[1]{>{\raggedleft\arraybackslash}p{#1}}
\DeclareSIUnit \parsec {pc}
\DeclareSIUnit \astrounit {au}

\providecommand{\orcit}[1]{\protect\href{https://orcid.org/#1}{\protect\includegraphics[width=8pt]{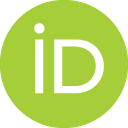}}}

\begin{document}
   \title{The SPACE Program. I. The featureless spectrum of \HD challenges sub-Neptune atmosphere trends}
   \author{
K. Angelique Kahle\orcit{0000-0001-7714-7551}\inst{1,2}\thanks{E-mail: kahle@mpia.de}
\and Jasmina Blecic\orcit{0000-0002-0769-9614}\inst{3,4}
\and Reza Ashtari\orcit{0000-0001-8943-9148}\inst{5}
\and Laura Kreidberg\orcit{0000-0003-0514-1147}\inst{1}
\and Yui Kawashima\orcit{0000-0003-3800-7518}\inst{6}
\and Patricio E. Cubillos\orcit{0000-0002-1347-2600}\inst{7,8}
\and Drake Deming\orcit{0000-0001-5727-4094}\inst{9}
\and James S. Jenkins\orcit{0000-0003-2733-8725}\inst{10,11}
\and Paul Molli\`ere\orcit{0000-0003-4096-7067}\inst{1}
\and Seth Redfield\orcit{0000-0003-3786-3486}\inst{12}
\and Qiushi Chris Tian\orcit{0009-0009-9148-2159}\inst{12,13}
\and Jose I. Vines\inst{14}
\and David J. Wilson\orcit{0000-0001-9667-9449}\inst{15}
\and Lorena Acuña\orcit{0000-0002-9147-7925}\inst{1}
\and Bertram Bitsch\orcit{0000-0002-8868-7649}\inst{16}
\and Jonathan Brande\orcit{0000-0002-2072-6541}\inst{17}
\and Kevin France\orcit{0000-0002-1002-3674}\inst{15}
\and Kevin B. Stevenson\orcit{0000-0002-7352-7941}\inst{5}
\and Ian J.M.\ Crossfield\inst{17,1}
\and Tansu Daylan\orcit{0000-0002-6939-9211}\inst{18,19}
\and Ian Dobbs-Dixon\inst{3,4}
\and Thomas M. Evans-Soma\orcit{0000-0001-5442-1300}\inst{20}
\and Cyril Gapp\orcit{0009-0007-9356-8576}\inst{1,2}
\and Antonio Garc\'ia Mu\~noz\orcit{0000-0003-1756-4825}\inst{21}
\and Kevin Heng\inst{22,23,24}
\and Renyu Hu\orcit{0000-0003-2215-8485}\inst{25,26}
\and Evgenya L. Shkolnik\orcit{0000-0002-7260-5821}\inst{27}
\and Keivan G. Stassun\orcit{0000-0002-3481-9052}\inst{28}
\and Johanna Teske\orcit{0009-0008-2801-5040}\inst{29, 30}
}
\institute{
\inst{1} Max Planck Institute for Astronomy, K\"{o}nigstuhl 17, 69117 Heidelberg, Germany \\
\inst{2} Department of Physics and Astronomy, Heidelberg University, Im Neuenheimer Feld 226, 69120 Heidelberg, Germany \\
\inst{3} Department of Physics, New York University Abu Dhabi, Abu Dhabi, UAE\\
\inst{4} Center for Astrophysics and Space Science (CASS), New York University Abu Dhabi, Abu Dhabi, UAE \\
\inst{5} JHU Applied Physics Laboratory, 11100 Johns Hopkins Rd, Laurel, MD 20723, USA \\
\inst{6} Department of Astronomy, Graduate School of Science, Kyoto University, Kyoto, Japan
 \\
\inst{7} Space Research Institute, Austrian Academy of Sciences, Schmiedlstrasse 6, A-8042, Graz, Austria \\
\inst{8} INAF -- Osservatorio Astroﬁsico di Torino, Via Osservatorio 20, 10025 Pino Torinese, Italy \\
\inst{9} University of Maryland: College Park, MD, US \\
\inst{10} Instituto de Estudios Astrof\'isicos, Facultad de Ingenier\'ia y Ciencias, Universidad Diego Portales, Av. Ej\'ercito 441, Santiago, Chile \\
\inst{11} Centro de Astrof\'isica y Tecnolog\'ias Afines (CATA), Casilla 36-D, Santiago, Chile \\
\inst{12} Astronomy Department and Van Vleck Observatory, Wesleyan University, Middletown, CT, USA \\
\inst{13} Department of Physics and Astronomy, The Johns Hopkins University, Baltimore, MD, USA \\
\inst{14} Instituto de Astronom\'ia, Universidad Cat\'olica del Norte, Angamos 0610, 1270709, Antofagasta, Chile \\
\inst{15} Laboratory for Atmospheric and Space Physics, University of Colorado, Boulder, Colorado, USA
 \\
\inst{16} Department of Physics, University College Cork, Cork, Ireland \\
\inst{17} Department of Physics and Astronomy, University of Kansas, 1082 Malott Hall, 1251 Wescoe Hall Dr, Lawrence, KS USA
 \\
\inst{18} Department of Physics, Washington University, St. Louis, MO 63130, USA \\
\inst{19} McDonnell Center for the Space Sciences, Washington University, St. Louis, MO 63130, USA \\
\inst{20} School of Information and Physical Sciences, University of Newcastle, Callaghan, NSW, Australia \\
\inst{21} Universit\'e Paris-Saclay, Universit\'e Paris Cit\'e, CEA, CNRS, AIM, 91191, Gif-sur-Yvette, France \\
\inst{22} Faculty of Physics, Ludwig Maximilian University, Munich, Germany \\
\inst{23} Department of Physics \& Astronomy, University College London, U.K. \\
\inst{24} Department of Physics, University of Warwick, Coventry, U.K. \\
\inst{25} Jet Propulsion Laboratory, California Institute of Technology, Pasadena, CA 91109, USA \\
\inst{26} Division of Geological and Planetary Sciences, California Institute of Technology, Pasadena, CA 91125, USA \\
\inst{27} School of Earth and Space Exploration, Arizona State University, Tempe, AZ 85281, USA \\
\inst{28} Department of Physics \& Astronomy, Vanderbilt University, Nashville, TN, USA \\
\inst{29} Earth and Planets Laboratory, Carnegie Institution for Science, Washington, DC, USA \\
\inst{30} Observatories, Carnegie Institution for Science, Pasadena, CA, USA \\}
   \titlerunning{The featureless spectrum of HD\,86226\,c}
   \authorrunning{K. A. Kahle et al.}
   \date{Received March 31, 2025; accepted July 13, 2025}
   \abstract
    {Sub-Neptune exoplanets are the most abundant type of planet known today. As they do not have a Solar System counterpart, many open questions exist about their composition and formation.
    Previous spectroscopic studies have ruled out aerosol-free hydrogen-helium dominated atmospheres for many characterized sub-Neptunes but are inconclusive about their exact atmospheric compositions. Here we characterize the hot (T$_\mathrm{eq}$=1311\,K) sub-Neptune HD\,86226\,c (R=2.2$\,\si{\rearth}$, M=$\SI{7.25}{\mearth}$), which orbits its G-type host star on a four-day orbit. The planet is located in a special part of the sub-Neptune parameter space: 
    Its high equilibrium temperature prohibits methane-based haze formation, increasing the chances for a clear atmosphere on this planet.
    We used Hubble Space Telescope data taken with WFC3 and STIS from the Sub-neptune Planetary Atmosphere Characterization Experiment (SPACE) Program to perform near-infrared ($1.1$--$1.7$\,\si{\micro\meter}) transmission spectroscopy and ultraviolet characterization of the host star.
    We report a featureless transmission spectrum that is consistent within $0.4\,\sigma$ with a constant transit depth of $418\pm14$\,ppm. The amplitude of this spectrum is only 0.01 scale heights for a H/He-dominated atmosphere, excluding a cloud-free solar-metallicity atmosphere on HD\,86226\,c with a confidence of $6.5\,\sigma$.
    Based on an atmospheric retrieval analysis and forward models of cloud and haze formation, we find that the featureless spectrum could be due to metal enrichment $[\mathrm{M}/\mathrm{H}]>2.3$ ($3\,\sigma$ confidence lower limit) of a cloudless atmosphere, or silicate (MgSiO$_3$), iron (Fe), or manganese sulfide (MnS) clouds. For these species, we performed a detailed investigation of cloud formation in high metallicity, high-temperature atmospheres. Our results highlight that HD\,86226\,c does not follow the aerosol trend of sub-Neptunes found by previous studies. Follow-up observations with the James Webb Space Telescope could determine whether this planet aligns with the recent detections of metal-enriched atmospheres or if it harbors a cloud species that is otherwise atypical for sub-Neptunes.} 
   \maketitle


\section{Introduction}\label{sec:intro}
Sub-Neptune exoplanets with a radius between $\SI{2}{\rearth}$ and $\SI{4}{\rearth}$ make up the most abundant planet class found by currently available surveys~\citep[e.g.,][]{Borucki2011kepler,Batalha2013kepler,Fulton2017gap}. The absence of a Solar System analog highlights the importance of examining sub-Neptune exoplanets to understand this planet class. However, observing sub-Neptunes is challenging with current technology, as these small-radius planets only cover a fraction of their host stars' light during their transit, leading to signal strengths of only a few hundred parts per million (ppm). Therefore, facilities with high sensitivities are needed to distinguish the planet signal from its star and various noise sources. The required precision is best met by space-based facilities such as the Hubble Space Telescope (HST) and the James Webb Space Telescope (JWST). But even these telescopes operate close to their limits to detect sub-Neptune planetary atmospheres.

The few sub-Neptune atmospheres characterized today hint at very diverse atmospheric conditions on these planets. Their compositions range from hydrogen-helium dominated atmospheres on GJ\,3470\,b~\citep[][]{Benneke2019lowzMie}, HD\,3167\,c~\citep[][]{Guilluy2021HD3167c}, K2-18\,b~\citep[][]{Madhusdan2023K218b}, and TOI-421\,b~\citep[][]{Davenport2025toi421} to metal-rich, miscible atmospheres as detected on TOI-270\,d~\citep[][]{Benneke2024TOI270d}. The molecular inventory differs across these planets: CH$_4$ is the dominant absorber in K2-18\,b and TOI-270\,d, H$_2$O has been seen in TOI-421\,b, and SO$_2$ absorption has been detected in GJ\,3470\,b~\citep[][]{Beatty2024So2_GJ3470b} and is thought to be a consequence of disequilibrium photochemistry. Furthermore, many of those atmospheres possibly host aerosols, as detected in GJ\,3470\,b~\citep[][]{Benneke2019lowzMie,Beatty2024So2_GJ3470b} and GJ\,1214\,b~\citep[][]{Kreidberg2014GJ1214b,Kempton2023gj1214bPhaseC}.

In contrast to the well-characterized planets, many of the observed sub-Neptune spectra are featureless or have muted spectral features. Examples of these include HD\,97658\,b~\citep[][]{Knutson2014hd97658b,Guo2020hd97658b}, HD\,106315\,b~\citep[][]{Kreidberg2022HD106315b}, TOI-836\,c~\citep[][]{Wallack2024compass}, GJ\,1214\,b~\citep[][]{Kreidberg2014GJ1214b,Kempton2023gj1214bPhaseC,Schlawin2024}, GJ\,9827\,d~\citep[][]{Roy2023waterworld,Piaulet-Ghorayeb2024_GJ9827d}, and GJ\,3090\,b~\citep{Ahrer2025gj3090b}, where muted features are detected for the last three planets. These muted spectral features were predicted as a consequence of high mean molecular weight atmospheres~\citep[e.g.,][]{Miller-Ricci_2009superearths} or high-altitude aerosols~\citep[e.g.,][]{Benneke2013cloudvsweight,Morley2013cloudsgj1214b}. Which of these effects dominates is still unclear for most of the observed planets. Trends in the spectra of Neptune-sized to sub-Neptune-sized planets suggest that there is a correlation between planet equilibrium temperature and atmospheric feature size~\citep[][]{Crossfield2017trend,Yu2021hazetrend,Brande2024neptune_trends}. \citet{Brande2024neptune_trends} attribute this trend to clouds, which rain out at low temperatures below $\SI{500}{\kelvin}$. For temperatures above $\SI{800}{\kelvin}$, a reduced formation of organic hazes may lead to clear atmospheres~\citep[e.g.,][]{Zahnle2009soot,Fortney2013framework,Crossfield2017trend,Gao2020CH4,Yu2021hazetrend}.
The characterized sample of sub-Neptunes is small, and not all planets follow this trend. More systematic observations are needed to fully understand which environment favors which atmospheric conditions.

\begin{figure}[t]
	\centering
	\includegraphics[width=0.49\textwidth]{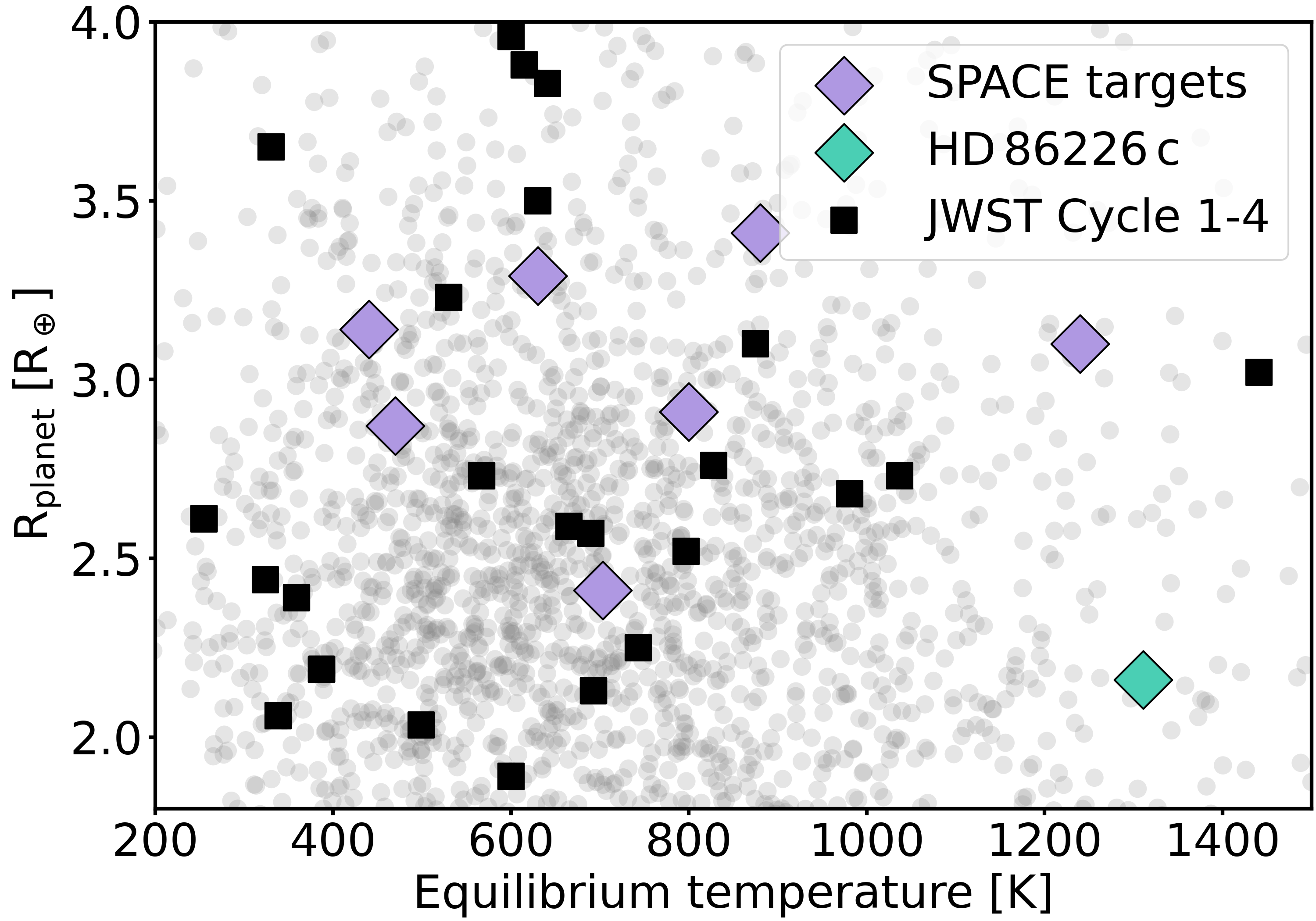}\\
    \includegraphics[width=0.49\textwidth]{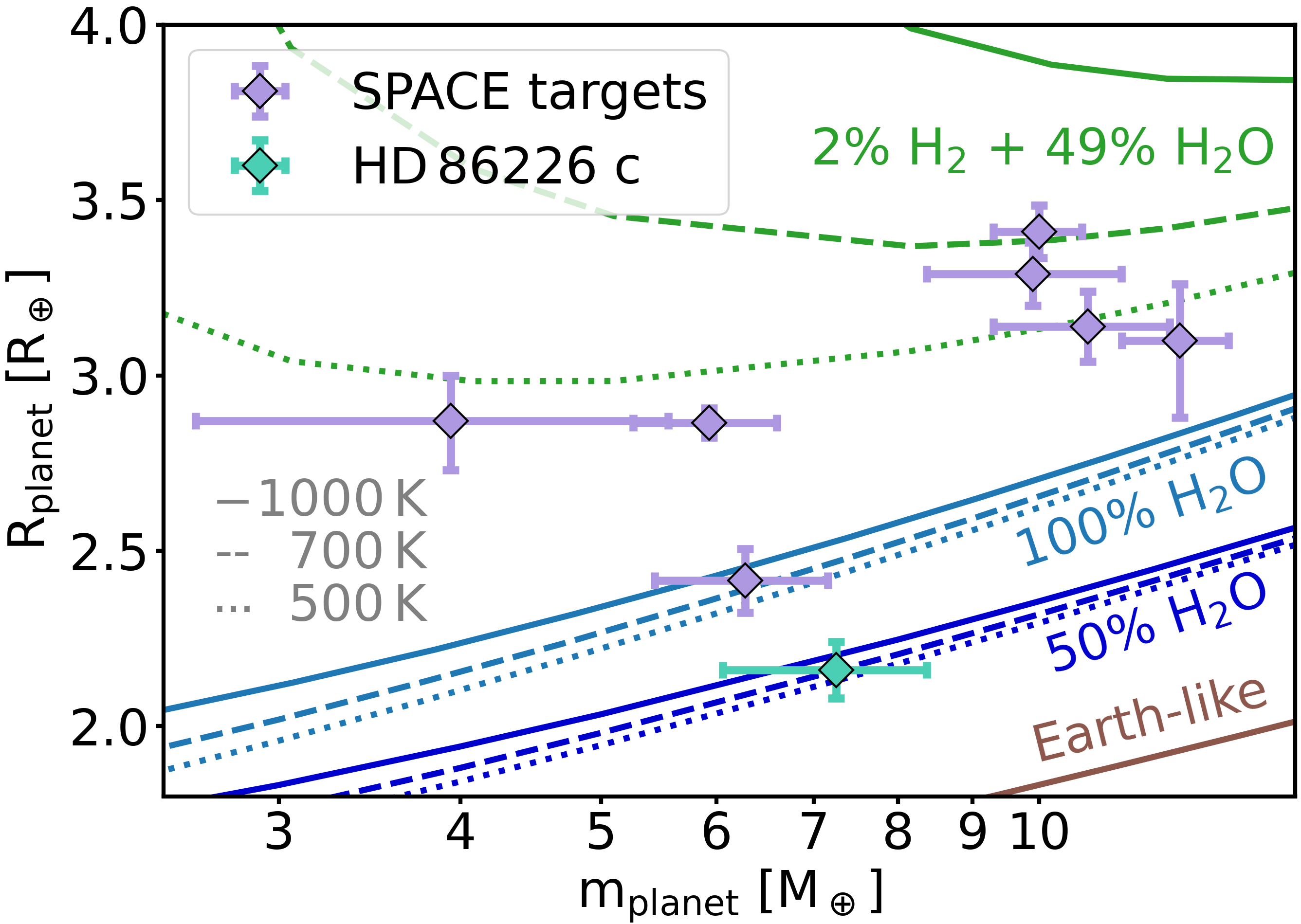}
	\caption[SPACE targets]{Sub-Neptune targets of the SPACE program. \HD is shown with a green square marker, and the other SPACE targets are shown in purple. Top: Temperature-radius plane. Gray circles show the known population of planets with radii between $\SI{1.8}{\rearth}$ and $\SI{4}{\rearth}$ based on the entries of the NASA Exoplanet Archive\footnotemark~in May 2024. Black crosses mark JWST Cycle 1-4 targets, except for the SPACE targets TOI-431\,d and \HD, which will be observed in Cycle 4. Bottom: Mass-radius plane. Color curves show models from~\citet{Zeng2019models} for various planetary compositions and temperatures.}
	\label{fig:intro:SPACE}
\end{figure}

The Sub-neptune Planetary Atmosphere Characterization Experiment (SPACE) with the HST is a multi-cycle treasury program that aims to conduct such a systematic survey~\citep{Kreidberg2022hst_prop}. It probes a sample of eight sub-Neptunes with near-infrared transmission spectroscopy using the Wide Field Camera 3 (WFC3\footnote{\url{https://www.stsci.edu/hst/instrumentation/wfc3}}) and combines these data with ultraviolet (UV) stellar characterization using the Space Telescope Imaging Spectrograph (STIS\footnote{\url{https://www.stsci.edu/hst/instrumentation/stis}}). As shown in the top panel of Fig.~\ref{fig:intro:SPACE}, the targets span a physically motivated grid across planet radius (2 -- 3.5$\,\si{\rearth}$) and temperature (300 -- 1400$\,\si{\kelvin}$), which allows to examine how the atmospheric conditions are altered with these parameters. Planet growth models~\citep{Zeng2019models} predict the presence of atmospheres on all SPACE targets based on their radii and masses between $\SI{3}{\mearth}$ and $\SI{13}{\mearth}$ (see bottom panel of Fig.~\ref{fig:intro:SPACE}). The associated host stars range in spectral type between M and G, corresponding to the types with the most planet detections today. The experiment aims to reveal how sub-Neptune atmospheres are shaped by metal enrichment, disequilibrium chemistry, and aerosols. Furthermore, it guides upcoming observations with JWST and identify sub-Neptunes with clear atmospheres.

\footnotetext{\url{https://exoplanetarchive.ipac.caltech.edu/}}

Here, we examine \HD, which is a $\SI{2.2}{\rearth}$ sub-Neptune that orbits its G-type host star on a $\SI{3.98}{\day}$ orbit. It has a high equilibrium temperature of $\SI{1310}{\kelvin}$~\citep{Teske2020HD86226c}, and its low mass of $\SI{7.25}{\mearth}$ suggests the presence of an atmosphere~\citep{Teske2020HD86226c} on the planet (see Fig.~\ref{fig:intro:SPACE}). These characteristics place \HD close to both the ``radius gap''~\citep{Fulton2017gap} and the ``hot Neptune desert''~\citep{Mazeh2016hotNeptuneDesert}. Moreover, \HD has a long-period giant planet companion, HD\,86226\,b~\citep[][]{Arriagada2010hd86226b}. These factors make \HD an interesting candidate for atmospheric characterization, as it could inform planet formation and evolution models. The G-type host star is both beneficial and hindering for the observations. While we do not expect high stellar activity based on its estimated age of $\SI{4.6}{\giga\year}$~\citep[][]{Teske2020HD86226c}, the expected transit depth is only on the order of $\SI{400}{\ppm}$ and  therefore just on the lower edge of what can be observed with HST. However, in this sparsely populated part of the temperature-radius parameter space, \HD is one of the best available targets for this study. 

The hot temperatures on \HD suggest a high chance for a clear atmosphere. At these temperatures, CO is the dominant carbon reservoir, instead of CH$_4$. This hinders the formation of organic hazes~\citep{Zahnle2009soot,Fortney2013framework,Gao2020CH4}, which could mute the observed atmospheric features. Previous observations of \HD with the Transiting Exoplanet Survey Satellite (TESS) and the $\SI{6.5}{\meter}$ Magellan Telescopes~\citep{Kokori2023Exoclock,Teske2020HD86226c} constrained its orbital parameters and mass. Possible volatile-rich atmospheric compositions were recently predicted by~\citet{Piette2023lavaworlds}, but no spectroscopic observations of the planet's atmosphere were conducted prior to this experiment.

Section~\ref{sec:observations} provides details on the observations of \HD and its host star. The independent data reductions of the transit spectra and stellar data are presented in Sect.~\ref{sec:analysis}. The resulting light curves are analyzed in Sect.~\ref{sec:LC}, where we also show the planetary spectrum. Forward models and retrievals of the planet spectrum are detailed and shown in Sect.~\ref{sec:models}. The implications of these results are discussed in Sect.~\ref{sec:discussion}, and we summarize our findings in Sect.~\ref{sec:summary}.

\section{Observations}\label{sec:observations}
\subsection{HST observations with WFC3 and STIS}
\HD was observed with HST as part of the Cycle 30 Large Treasury Program 17192 (PI: Laura Kreidberg). For the transit observations, we used WFC3 with the G141 grism, spanning the wavelength between $1.1$ and $\SI{1.7}{\micro\meter}$. Individual exposures of the time series observations used the MULTIACCUM mode with three nondestructive readouts and a total integration time of $\SI{46.696}{\second}$. The spatial scanning mode was used to allow for longer integration times before saturation by spreading the light across the detector~\citep[see e.g.,][]{Deming2013WFC3}. The scan rate of 0.754\,''s$^{-1}$ yields a scan height of 35.2'' for the final readout, corresponding to approximately 290 pixels on the detector.

Roundtrip spatial scans (alternating between forward and reverse) were used to maximize the observing time per orbit, leading to 26 exposures observed per HST orbit. In addition, a direct image of HD\,86226 was obtained at the start of each orbit using the F139M filter and an exposure time of $\SI{5.118}{\second}$. We achieved an observing efficiency of 42\%, including overheads and target acquisition time. 

Nine transits were observed, with four HST orbits per transit between May 2023 and January 2025. A guiding failure occurred on 2024 January 10, when the fine guidance sensor could only acquire one guiding star. The telescope observed orbit 29 on gyro control, which resulted in the spectrum moving by up to one pixel row on the detector. As the flux measured in this dataset is correlated with the position of the spectral trace on the detector (see Sect.~\ref{subsec:pacmam}), the uncertainty of the flux in orbit 29 is increased. However, the data of this orbit generally remain usable for transit spectroscopy. 

The host star HD\,86226 was also observed with STIS over two orbits in May 2023. These observations consist of three exposures using the G140M, G140L and E230M gratings, covering 1194--1248\,\AA\ (including the \ion{H}{i} Lyman\,$\alpha$ 1215\,\AA\ line),  1160--1715\,\AA\, and 2274--3118\,\AA\ respectively, with exposure times of $\SI{1803.169}{\second}$, $\SI{1978.177}{\second}$ and $\SI{200.020}{\second}$.

\subsection{Host star monitoring}
In addition to HST spectroscopy, we also conducted long-term photometric monitoring of the host star HD\,86226 from the ground to verify the level of stellar activity. On more than 60 observing nights from March 2023 to April 2024, HD\,86226 was observed with the automated 24-inch PlaneWave CDK telescope at Wesleyan University's Van Vleck Observatory in Connecticut, USA. The telescope has an aperture of $0.61 \,\mathrm{m}$ and is equipped with a FLI PL4240 camera. The field of view is approximately $23\,\mathrm{arcmin}\,\times\,23\,\mathrm{arcmin}$ at approximately $0.7 \,\mathrm{arcsec}$ per pixel. Typically, twenty exposures were taken every observing night in each of the BVRI bands, with exposure times of $6.0 \, \mathrm{s}$, $5.0 \,\mathrm{s}$, $2.0 \,\mathrm{s}$, and $1.5 \,\mathrm{s}$, respectively. We present the reduction of this dataset and the photometric monitoring results in Sect.~\ref{subsec:monitoring-reduction}.

\section{Data reduction and analysis}\label{sec:analysis}
We reduced the HST/WFC3 data with two independent pipelines: \texttt{PACMAN}~\citep{Zieba2022Pacman} and \texttt{Eureka!}~\citep{Bell2022Eureka}. Sect.~\ref{subsec:pacmam} covers the data reduction with \texttt{PACMAN}, a well-tested~\citep[e.g.,][]{Kreidberg2014GJ1214b,Kreidberg2014wasp43b,kreidberg2018wasp103b,kreidberg2018wasp107b,Colon2020Akelt11b} pipeline optimized for HST/WFC3 data. Details about the  \texttt{Eureka!} analysis can be found in Section~\ref{subsec:eureka}. The stellar parameters used in these analyses are estimated with \texttt{ARIADNE} in Sect.\ref{subsec:ariadne}. Section~\ref{subsec:star_reduction} describes the reduction of the STIS data of HD\,86226, and the monitoring results can be found in Sect.~\ref{subsec:monitoring-reduction}.

\subsection{Transit reduction with \texttt{PACMAN}}\label{subsec:pacmam}
The \texttt{PACMAN} reduction used the \texttt{\_ima} files provided by the Barbara A. Mikulski Archive for Space Telescopes (MAST\footnote{\url{https://mast.stsci.edu/portal/Mashup/Clients/Mast/Portal.html}}) as a starting point. The time stamps of these files were corrected for the motion of the HST with respect to the Solar System Barycenter.

The spectrum for each exposure was derived from the three non-destructive up-the-ramp readouts using an optimal extraction algorithm~\citep{Horne1986optextr}. To extract each spectrum, \texttt{PACMAN} created a sequence of differenced images consisting of the difference between subsequent up-the-ramp samples. The differenced images contain the first order of the spectral trace (columns 153 to 334). Each differenced image was background corrected by subtracting the median value of all pixels with a value below 500 counts. The optimal extraction algorithm was applied to all rows in the differenced image within 40 rows of the edges of the spectral trace in the spatial direction. These edges were determined by accumulating the flux in the spatial direction and identifying the rows between which the flux declines the strongest. The final spectrum and its uncertainty per exposure were determined by summing the spectra and variances for all of the differenced images.

For wavelength calibration, we used the stellar model grids of \citet{Castelli2003_stellarmodel}. The model assumes the stellar parameters $\log(\mathrm{Z}) = -0.07^{+0.15}_{-0.14}$ and $\log_{10}(\mathrm{g[\mathrm{cm\,s}^{-2}]})=4.55^{+0.40}_{-0.48}$, obtained from fits to the stellar spectral energy distribution (SED) with \texttt{ARIADNE}, which are described in Sect.~\ref{subsec:ariadne}. To obtain the calibration, we interpolated the stellar model and fitted it to the extracted observed spectrum. In this fit, we derived the shift in wavelength direction and allowed for a linear stretch in flux and wavelength.

\begin{figure}[t]
	\centering
	\includegraphics[width=0.49\textwidth]{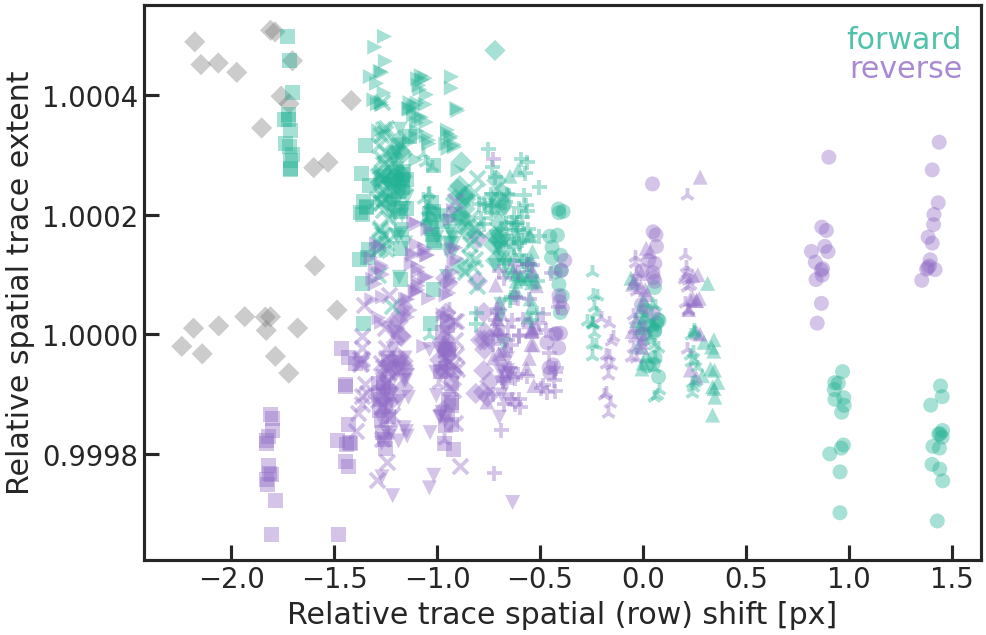}
	\caption[Spectrum shift on detector]{Relative stretch of the light trace as a function of trace position on the WFC3 detector. Both values are relative to the first exposure of the observations. Values for forward and reverse scan directions are green and purple, respectively. Different symbols indicate the nine individual visits of the planet. The gray diamonds mark orbit 29, which was removed for the \texttt{PACMAN} analysis.}
	\label{fig:red:rowshift_stretch}
\end{figure}

The raw \HD light curve shows a non-constant offset between spatial scans in the forward and reverse direction. While the known upstream-downstream effect~\citep{McCullough2012wfc3updown} more typically results in a constant offset between scan directions, a similar non-constant offset was also seen in WFC3 observations of 55\,Cancri\,e~\citep{Tsiaras2015_55cancrie}. This behavior might be a consequence of the fast scan speed used for observing bright objects: For fast scans, the position of the spectrum in the spatial direction can shift by up to two pixels between orbits (see Fig~\ref{fig:red:rowshift_stretch}). Because the detector is read out row-by-row and has no shutter, a shift in spatial position effectively changes the exposure time. This introduces a systematic change in flux with the position of the trace in the spatial (row) direction on the detector. 

To correct this effect, we introduced a new systematics decorrelation parameter, the ``row shift'', based on the position of the spectral trace.
We independently assumed a linear correlation between flux and shift of spectral trace in spatial direction (row shift) for both scan directions. The row shift was determined for each exposure by summing the flux of the first non-destructive readout along the dispersion direction (columns) and comparing this spatial profile to the profile of the first exposure of all observations. This resulted in the row shift relative to the first exposure, which was used in the further analysis of the data.

Figure~\ref{fig:red:rowshift_stretch} shows that the spatial extent of the spectral trace on the detector is correlated with the position of the trace. Analogous to the row shift measurement, the relative trace extent was derived from the spatial profiles of the last detector readouts of each exposure. The profile of each exposure was compared to the profile of the first exposure of the observations by fitting a shift and a linear stretch in the spatial direction. While a tight negative linear correlation is apparent for scans in the forward direction, the trace extents differ stronger for scans in the reverse direction and show a change in slope at $\SI{-0.2}{\pixel}$. To simplify the correction of this systematic, we continued assuming a single linear correlation for each scan direction.

The method of determining the relative trace extent for each exposure is inaccurate, as spatial profiles of individual exposures can deviate in shape. In these cases, the fit is not only guided by the trace extent but also by other characteristics of the spatial profile. Therefore, we directly de-correlated the flux against the row shift derived from each first non-destructive exposure during the light curve fits in Sect.~\ref{subsec:LCwithPACMAn}.

\subsection{Data reduction with \texttt{Eureka!}}\label{subsec:eureka}
\texttt{Eureka$!$} \citep{Bell2022} is a publicly available, community-developed pipeline designed to reduce and analyze exoplanet time-series observations with particular emphasis on data from JWST. It converts raw, uncalibrated FITS images into precise exoplanet transmission or emission spectra. It features a modular design of six stages, of which four (Stages 3--6) are employed for HST WFC3 transit observations.

Each stage in the pipeline is guided by ``\texttt{Eureka$!$} Control Files'' (ECFs) and ``\texttt{Eureka$!$} Parameter Files'' (EPFs), with the latter used to adjust transit model fit parameters. \texttt{Eureka!} currently provides template ECFs for JWST’s MIRI, NIRCam, NIRSpec instruments, and HST’s WFC3 instrument. Further details can be found on the \texttt{Eureka$!$} \href{https://eurekadocs.readthedocs.io/en/latest/index.html}{ReadTheDocs} page.

To reduce HST WFC3 data with \texttt{Eureka!}, we used the \texttt{\_ima} files provided by MAST as a starting point. Stage 3 of the pipeline performed background subtraction and optimal spectral extraction on calibrated image data~\citep{Bell2022, Ashtari2024}. The extraction apertures and regions for background estimation were assigned individually for each visit to achieve the best possible spectrum extraction. As already demonstrated with the \texttt{PACMAN} reduction in Fig.~\ref{fig:red:rowshift_stretch}, this was necessary due to the non-negligible shift of the spectral trace in scan-direction. This stage resulted in a time series of 1D spectra. 

Stage 4 used this time series of 1D spectra to generate the light curve by binning it along the wavelength axis, then removed drift and jitter to produce spectroscopic light curves~\citep{Ashtari2024}. Stages 5 and 6 of \texttt{Eureka!} conducted the light curve fitting and plotting of the results. Details about the light curve fitting can be found in Sect.~\ref{subsec:LCwithEureka}.

\texttt{Eureka!} also detected the non-negligible shift of the spectral trace in the scan direction on the detector (see Fig.~\ref{fig:red:rowshift_stretch}). For seven of the eight transits analyzed with \texttt{Eureka!}, the shift in scan direction is greater than the recommended 0.2 pixel (peak-to-peak) threshold for precise time series observations \citep{Stevenson2019}. Because each pixel has distinct sensitivity characteristics, spatial shifting of the spectrum introduces variability in the measured flux \citep{Stevenson2019}. Correcting for these effects requires precise modeling of positional shifts that are dynamic and do not follow simple patterns. As such, modeling is currently not implemented in \texttt{Eureka!}, the positional dependence limits the precision that can be achieved with this pipeline. The limitations and capabilities of \texttt{Eureka!} for fitting the light curves of this dataset is discussed further in Sects.~\ref{subsec:LCwithEureka} and~\ref{subsec:PACMANandEureka}.

\begin{table}[]
    \centering
    \caption{Stellar parameters of HD\,86226.}
    \begin{tabular}{lr}\hline \hline
         Parameter &  Value \\ \hline
         $T_{\rm eff}$ [K] & 5980$\pm$70  \\
         $\log_{10}(\mathrm{g[\mathrm{cm\,s}^{-2}]})$ & $4.55^{+0.40}_{-0.48}$ \\
         $[\mathrm{Fe/H}]$ [dex] & $-0.07^{+0.15}_{-0.14}$ \\
         A$_{\rm V}$ [mag] & 0.06$\pm$0.03\\ 
         R [R$_\odot$] & 1.05 $\pm$ 0.02\\
         d [pc] & 45.49* \\
         \hline    
    \end{tabular}
    \tablefoot{The stellar parameters were derived using \texttt{ARIADNE}. An exception is the distance (marked with *), which was adapted from the Gaia data release 3~\citep{Gaia2016mission,Gaia2023DR3}. }
    \label{tab:ariadne}
\end{table}

\subsection{Stellar parameter estimation with \texttt{ARIADNE}}\label{subsec:ariadne}
For the analysis of the planet data, it is essential to obtain precise parameters of the star, especially those needed to determine the best-fitting stellar models. We derived the stellar parameters based on fits to the SED with \texttt{ARIADNE}~\citep{Vines2022ARIADNE} code. 

\texttt{ARIADNE} is an open-source Python package designed to automatically fit the SED of stars to different stellar atmosphere model grids. For HD\,86226, we used the Phoenix v2 \citep{Husser2013}, BT-Settl \citep{Hauschildt1999, Allard2012}, \cite{Kurucz}, and \cite{Castelli2003_stellarmodel} model grids, which were previously convolved with several broadband filters. \texttt{ARIADNE} operates under a Bayesian framework, where each grid is modeled using a Nested Sampling \citep{nestedsampling1, nestedsampling2, dynamicnestedsampling} algorithm as implemented by \texttt{dynesty} \citep{dynesty}. The algorithm calculates the evidence of each model, while also computing the posterior distributions for the effective temperature, $\log_{10}(g)$, [Fe/H], extinction in V, and radius of the star.
After each grid was fitted, we combined the posterior distributions using Bayesian model averaging, which leverages the evidence of each model and accounts for model intrinsic uncertainties. We summarize the results from \texttt{ARIADNE} in Table \ref{tab:ariadne}.

\subsection{STIS stellar spectral analysis}\label{subsec:star_reduction}
\begin{figure}
    \centering
    \includegraphics[trim = 5mm 5mm 5mm 5mm,clip, width=\linewidth]{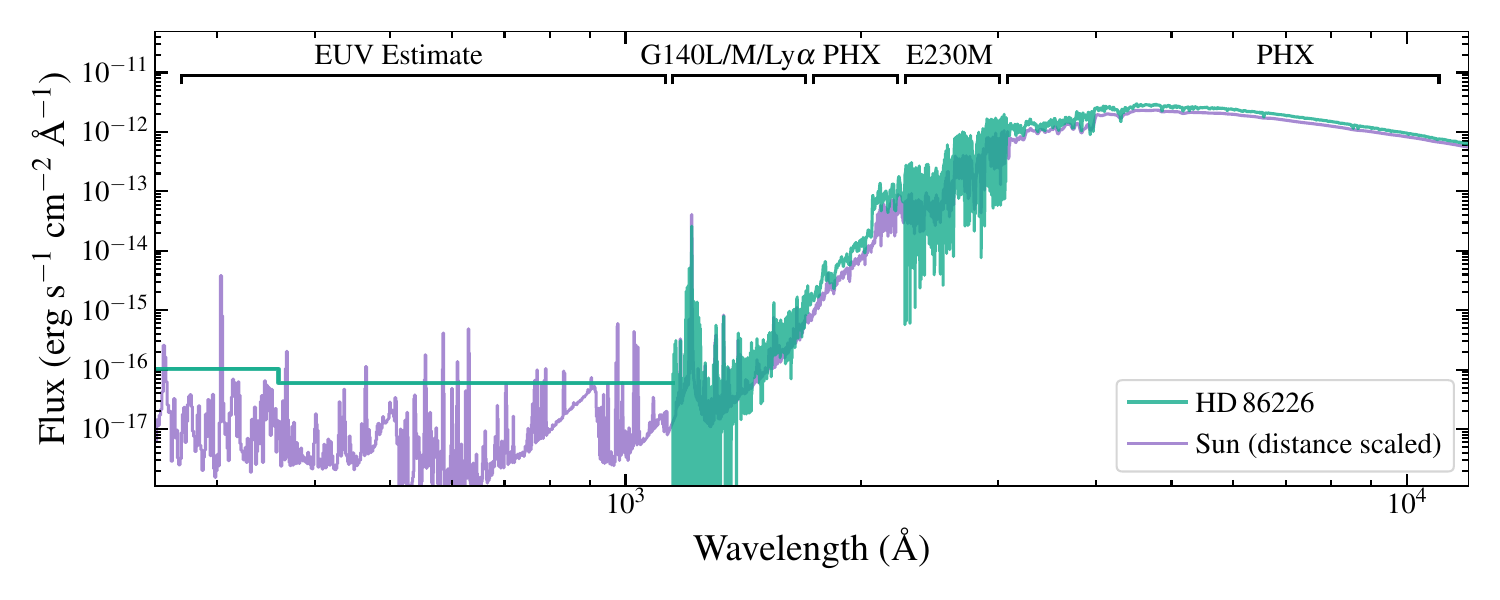}
    \caption{Full reconstructed SED of HD\,86226. Various components ar described in the text labeled in the top legend. The solar spectrum was taken from \cite{woodsetal09-1} and scaled to the distance of HD\,86226. The Phoenix models of HD\,86226 were binned to a lower resolution for this figure to allow for better comparison to the solar spectrum.}
    \label{fig:red:sed}
\end{figure}
The STIS spectra were automatically calibrated with {\sc CalSTIS v.3.4.2} and retrieved from MAST. The automated extraction successfully identified the spectral trace in all three gratings, and no custom extractions were required. We detect multiple strong emission lines in the spectra and a clear continuum signal redward of $\approx$1350$\,\AA$. To splice the three spectra together, as well as the orders of the echelle E230M spectra, we retrieved the Hubble Advanced Spectral Products (HASP\footnote{\url{https://archive.stsci.edu/missions-and-data/hst/hasp}}) complete coadd of the spectrum.

The Lyman\,$\alpha$ line is by far the brightest far-UV (FUV) line, but is strongly absorbed by the local interstellar medium~\citep[LISM, see][]{Redfield2008lism} and must be reconstructed in order to complete the FUV spectrum. Fortunately, the wings of the Lyman\,$\alpha$ line are clearly detected in the G140M spectrum, allowing us to reconstruct the intrinsic line profile with \texttt {lyapy}\footnote{\url{https://github.com/allisony/lyapy}}\citep{youngbloodetal16-1, youngbloodetal22-1}. We used the example start file provided with the package, changing the input spectrum, line spread function (to the G140M LSF), and the number of steps to 50000. We also fixed the self-reversal parameter $p=2.4$~\citep{Taylor2024stellarparameters}. The key parameters of the Lyman\,$\alpha$ reconstruction are given in Table \ref{tab:red:lya}. The LISM column density and velocity are consistent with the sample of LISM absorptions of nearby stars~\citep{Redfield2008lism} within $3\,\sigma$. We note that the large uncertainties in the derived values are caused by an airglow line at the position where we expect the LISM absorption. This caused small uncertainties in the background subtraction to propagate into large
uncertainties in the LISM constraints.

\begin{table}[]
    \centering
    \caption{Key parameters of the reconstructed Lyman\,$\alpha$ line.}
    \begin{tabular}{lr}\hline \hline
         Parameter &  Value \\ \hline
         $F_{\mathrm{Ly}\alpha}$ [$\mathrm{erg}\,\mathrm{s}^{-1}\,\mathrm{cm}^{-2}$] & $1.18^{+0.09}_{-0.08} \times10^{-14}$\\
         $\log_{10}$(N(\ion{H}{i}[cm$^{-2}$]) & $18^{+0.2}_{-0.3}$\\
         $v_{\mathrm{Ly}\alpha}$ [km\,s$^{-1}$] & $6.5\pm12$\\
         
         $v_{\mathrm{LISM}}$ [km\,s$^{-1}$] & $-28^{+10}_{-11}$\\ \hline
    
    \end{tabular}
    \tablefoot{The integrated flux of the intrinsic line, LISM hydrogen column density, and the radial velocities of the line and the LISM.}
    \label{tab:red:lya}
\end{table}

The gap between the G140L and G230M spectra covering 1715--2274\,\AA\ and wavelengths greater than 3118\,\AA\ were filled with a Phoenix model spectrum from the Lyon BT-Settl CIFIST~2011\_2015 grid \citep{allard16-1, baraffeetal15-1} using the stellar parameters from Sect.~\ref{subsec:ariadne}. Uncertainties were estimated using models with that account for the uncertainty in $T_{\mathrm{eff}}$. We find that the model is a good match to both sides of the gap in the UV spectrum, although the model diverges from the spectrum at wavelengths smaller than $1700$\,\AA. This indicates that the continuum contribution from the stellar chromosphere begins to dominate over the photosphere flux at those wavelengths. 

Wavelengths below 1160\,\AA\ were filled using the FUV to extreme UV (EUV) scaling relations from \cite{franceetal18-1}. We used the scaling relationship for the \ion{Si}{iv}\,1400\,\AA\ doublet as it is detected with high S/N in the G140L spectrum. The line flux was measured by integrating the G140L spectrum over the region 1390--1410\,\AA\ then subtracting a flat continuum flux fitted to the surrounding spectrum. We find $F_{\mathrm{\ion{Si}{iv}}} = (1.25\pm0.18) \times 10^{-15}\,\mathrm{erg}\,\mathrm{s}^{-1}\,\mathrm{cm}^{-2}$. Propagating both the error on the flux and on the scaling relationships, we find $\log_{10}(F (90-360\,\mathrm{\AA}) [\mathrm{erg}\,\mathrm{s}^{-1}\,\mathrm{cm}^{-2}]) = -13.6^{+1.00}_{-0.04}$ and $\log_{10}(F (360-911\,\mathrm{\AA}) [\mathrm{erg}\,\mathrm{s}^{-1}\,\mathrm{cm}^{-2}]) = -13.8^{+1.00}_{-0.04}$. We generated a two-step spectrum for the region 90-1160\,\AA\ with these relationships, extending the higher wavelength band to meet the blue end of the STIS spectrum. 

Our completed SED is shown in Fig.~\ref{fig:red:sed} with the various components labeled. We compare the SED to that of the Sun. As might be expected from the stellar parameters, they are almost identical. The slight offsets between the spectra are within the uncertainties of the reconstructed spectrum. 

The flux at wavelengths shorter than 1400\,\AA\ is strongly affected by stellar activity, in addition to the effective temperature. The close agreement of HD\,86226 with the flux of the Sun therefore implies similar activity levels of both stars. This is also traced by the 25-day rotational period of HD\,86226~\citep{Arriagada2011Prot_star}, which is similar to the solar rotation period at the equator.

\subsection{Host star optical monitoring} \label{subsec:monitoring-reduction}
The ground-based photometric monitoring data were handled and reduced with a customized pipeline. The World Coordinate System (WCS) parameters of the images were determined by the Astrometric STAcking Program \texttt{ASTAP}\footnote{\url{https://www.hnsky.org/astap.htm}}, using star databases made with Gaia data releases. We used Astropy's \texttt{ccdproc} package \citep{Craig2015ccdproc} for bias, dark, and flat calibrations and \texttt{photutils}~\citep{larry_bradley_2024_photutils} for aperture photometry. The aperture photometry process found centroids of stars in the field and placed circular apertures with a radius of $1.4$ times the smallest full width at half maximum (FWHM) of stars in the field. Local backgrounds were determined with annular apertures between $2.4$ and $3.6$ times the FWHM and subtracted. We used field stars and their magnitudes obtained from SIMBAD\footnote{\url{https://simbad.cds.unistra.fr/simbad/}} as standards, with seven stars in the B band, five in the V band, and two in the R and I bands each.

\begin{figure}[]
    \centering
    \includegraphics[width=0.49\textwidth]{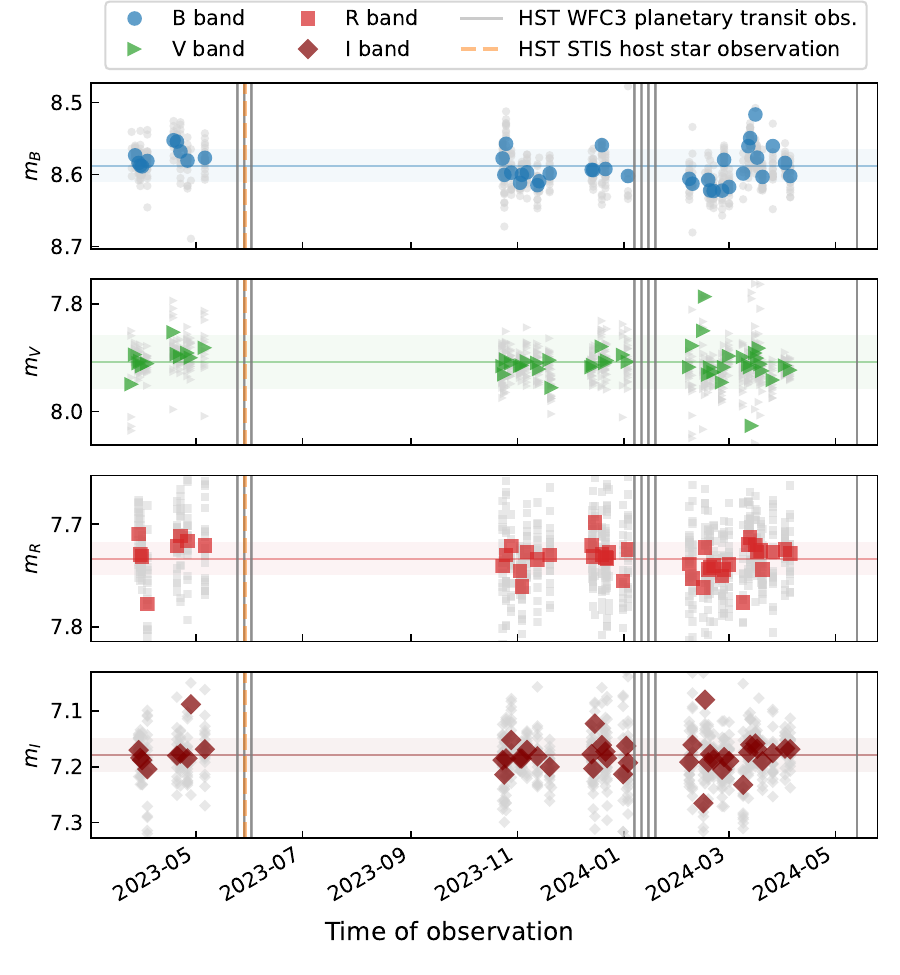}
    \caption{Ground-based photometric monitoring light curve of HD\,86226. The large colored markers represent nightly mean magnitudes. The small gray markers represent individual exposures. The sigma-clipped mean magnitudes averaging the entire light curves and the $\pm 1\,\sigma_\mathrm{all}$ ranges are represented with horizontal lines and shaded regions. Vertical lines correspond to HST observations; the last HST visit in January 2025 is not shown. We identify no evidence of stellar activity during the monitoring period.}
    \label{fig:monitoring-light-curve}
\end{figure}

The light curves shown in Fig.~\ref{fig:monitoring-light-curve} retain the magnitude of HD\,86226 as measured with single exposures and binned by observing nights. The figure also highlights the HST visit times. The photometric monitoring dataset covers most HST visits except for the last two WFC3 observations. We characterize the statistics of the photometry results in Table~\ref{tab:monitoring-result}. The sigma-clipped mean magnitudes, averaging entire light curves, match well with literature values, indicating good overall photometry quality (see Cols.~2 and 5 in Table~\ref{tab:monitoring-result}). The night-to-night standard deviation is comparable to the median exposure-to-exposure standard deviation within given nights (Cols.~3 and 4 in Table~\ref{tab:monitoring-result}), indicating no night-to-night brightness fluctuation. The lack of significant photometric stellar activity matches our expectation that the host star is relatively quiet and its activity should not significantly impact the transmission spectrum.

\begin{table}[]
    \centering
    \caption{HD\,86226 ground-based photometry.}
    \label{tab:monitoring-result}
    \begin{tabular}{C{0.5cm}cccc}
        \hline
        \hline
        Band & Mean mag. & $\sigma_\mathrm{all}$ & $\mathrm{med}(\sigma_\mathrm{exp})$ & Lit. mag. \\
         & (mag) & (mag) & (mag) & (mag) \\
        \hline
        B & $8.588$ & $0.023$ & $0.015$ & $8.56 \pm 0.02$ (a) \\
        V & $7.908$ & $0.031$ & $0.019$ & $7.93 \pm 0.01$ (a) \\
        R & $7.734$ & $0.016$ & $0.028$ & $7.71 \pm 0.11$ (b) \\
        I & $7.179$ & $0.030$ & $0.033$ & N/A \\
        \hline
    \end{tabular}
    \tablefoot{For each band, we report the sigma-clipped mean magnitude averaging the entire light curve in Col.~2 and the night-to-night standard deviation $\sigma_\mathrm{all}$ in Col.~3. $\sigma_\mathrm{exp}$ notes the exposure-to-exposure standard deviation in a given night, whose median value across all nights $\mathrm{med}(\sigma_\mathrm{exp})$ is reported in Col.~4. The last column lists the magnitude of HD\,86226 in the literature: (a) \cite{monitoring-a}, (b) \cite{monitoring-b}.}
\end{table}

\section{Light curve fitting}\label{sec:LC}
\subsection{Light curve fitting with \texttt{PACMAN}}
\label{subsec:LCwithPACMAn}
We fitted the \texttt{PACMAN}-reduced data simultaneously with a \texttt{batman}~\citep{Kreidberg2015batman} transit model and a systematics model, which are applied multiplicatively. Following common practice, we removed the first orbit in each visit due to its relatively strong exponential ramp. As the flux of the first exposure in each orbit is much lower than the later exposures, we also removed the first exposure of each orbit. Lastly, we removed all exposures from orbit 29, as the increased spectral trace instability hindered the row shift correction (see Sect.~\ref{sec:observations} and Fig.~\ref{fig:red:rowshift_stretch}).

For the transit model, we fixed the planet's semi-major axis, inclination, eccentricity, and argument of periastron to the values derived by \citet{Teske2020HD86226c}. The period of the planet was fixed to the value derived by \citet{Kokori2023Exoclock}. We used a quadratic limb darkening law and fixed the limb darkening coefficients to the values obtained from the \citet{Castelli2003_stellarmodel} stellar grid with the \texttt{exotic-LD}~\citep{Grant2024ExoticLD} Python package. Attempts to fit the limb darkening coefficients with a linear limb darkening law and free parameters led to fit coefficients similar to the model values. In contrast, attempts to fit a quadratic law with free parameters resulted in strong degeneracies between the free parameters. Due to this, fixing the limb darkening coefficients to the model values for a quadratic law provided the most accurate solution for our fit. Therefore, free parameters for the astrophysical model were only the transit mid-time and the transit depth. For both parameters, we fitted one common value across all visits. The transit mid-time was hereby extrapolated from the first to the subsequent visits using the orbital period of the planet.

For the systematics model, we fitted a constant flux and offset between forward and reverse scan flux for each visit. In addition, the flux behavior with time for each individual visit was described by a linear slope across the orbits and an exponential ramp with the same shape for each orbit. The exponential ramp was parametrized by 
\begin{equation}
    \exp(-r_1 \times t_\mathrm{orbit} - r_2),
\end{equation}
where $t_\mathrm{orbit}$ is the time since the start of the associated orbit and $r_1$ and $r_2$ are free parameters. In addition, a linear slope in flux with row shift (see Sect.~\ref{subsec:pacmam}) was fit to all visits simultaneously for forward and reverse scans, respectively. To account for possible neglected sources of uncertainty, we additionally fitted a scale factor that inflates the uncertainties to achieve a reduced $\chi^2$ of one.

We ran a Marcov Chain Monte Carlo (MCMC) algorithm with the Python package \texttt{emcee}~\citep{ForemanMackey2013emcee} to obtain the final parameter estimates and uncertainties. We used broad, uninformed linear priors for all parameters of the systematics model and the transit depth. We ran chains with 100000 steps and 500 walkers, removing the first 4000 steps of each chain as burn-in. This ensured that the chains were longer than 50 times the autocorrelation time for astrophysical and systemic parameters and that the chains converged. The uncertainties on the parameters were derived from the posterior distributions and based on the Gaussian $1\,\sigma$ deviations of the parameters from their median value. Minor correlations are seen in the posterior distributions of the constant flux offsets and the linear slopes of flux with time. Furthermore, the row shift slopes between forward and reverse scans are strongly correlated. Since both parameters describe the shift of the spectral trace on the detector, a tight correlation between these two systemic corrections is expected. The obtained transit depth is not strongly correlated with the other parameters. 

\begin{figure*}[t]
	\centering
	\includegraphics[width=0.99\textwidth]{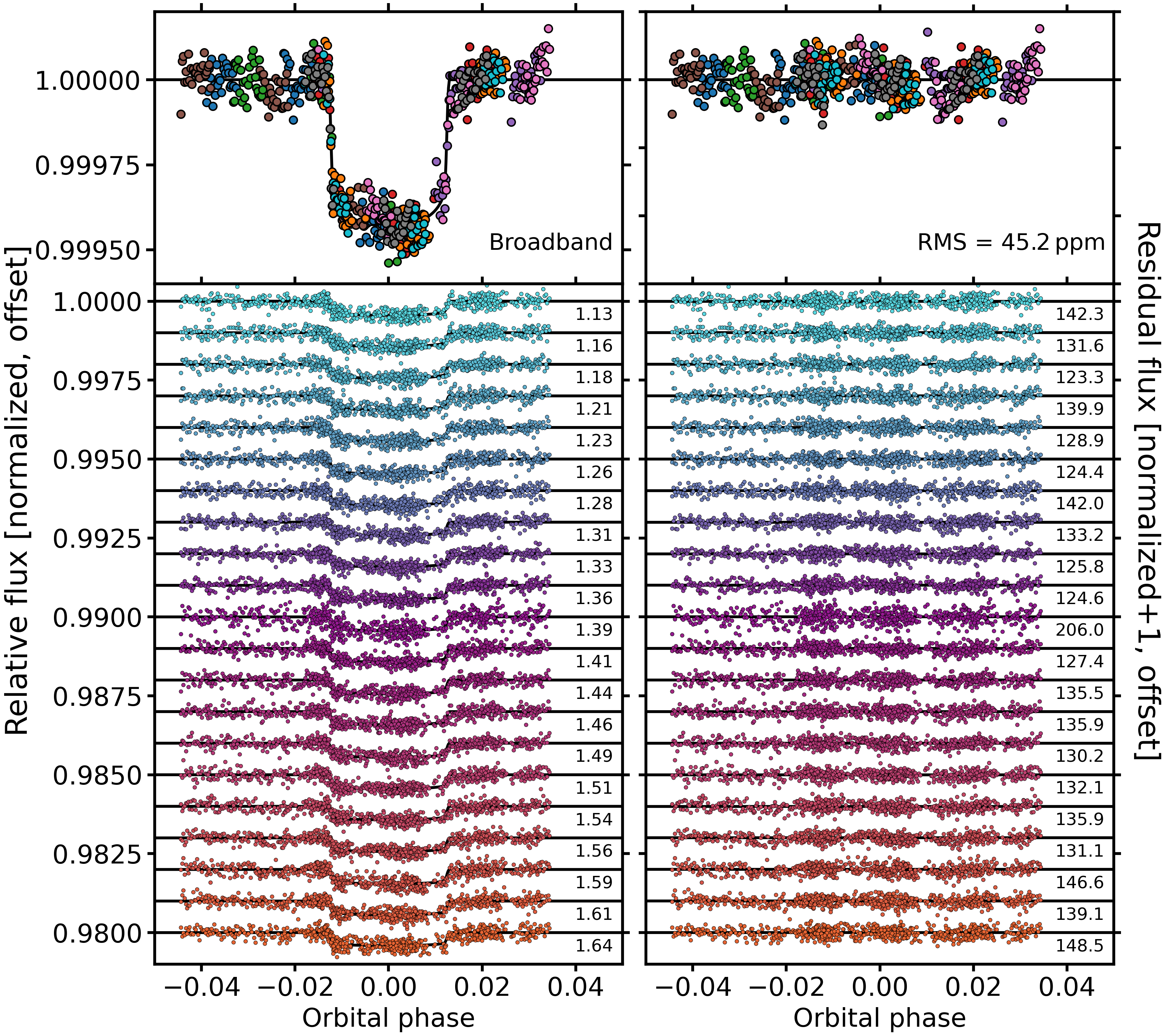}
	\caption{Broadband (top panel) and spectroscopic (bottom panel) light curves obtained with \texttt{PACMAN} by combining all nine observed transits of \HD. Differently colored data points in the broadband light curve mark the respective visits. The right panel shows the residuals from the fit transit model. Numbers in the bottom panel state the wavelength bin in $\mu$m (left panel) and the RMS in ppm (right panel).}
	\label{fig:red:pacman_white_lc}
\end{figure*}

The broadband light curve is shown in Fig.~\ref{fig:red:pacman_white_lc}. The root mean square (RMS) noise of 45.2\,ppm is above the expected photon noise of 25.2\,ppm, but within the best achieved RMS values for HST data~\citep[see e.g.,][]{Kreidberg2014GJ1214b, Tsiaras2015_55cancrie}. Major contributors to the remaining red noise are a wave-like residual pattern with time and the systematically lower egress flux. The wave-like red noise can be removed by fitting a second-order polynomial with time to the flux of each visit instead of a linear slope. We tested if this model is a better fit for the systematics by calculating the Bayesian information criterion~\citep[BIC,][]{Kass1995bayes} for both models. Using a second-order polynomial improved the BIC of the broadband light curve fit significantly by $78.4$, as compared to the linear slope with time. However, as seen in Table~\ref{tab:red:BIC}, this improvement only holds for five spectroscopic bins, while the first-order polynomial is preferred for most bins. To avoid overfitting the spectroscopic data, we opted to fit a linear slope with time to all datasets. 

The observed egress flux is systematically lower than the model flux by approximately 50\,ppm. Since this lower flux is not seen during ingress, we can exclude issues with the limb darkening as the cause of this lower egress flux. Most of this offset can be accounted for by fitting a second-order flux baseline for each visit. However, due to the many steps involved in fitting the light curve, we cannot unambiguously say if the decreased egress flux is astrophysical or just an artifact of the fitting process. A continuous time series observation of this planet could clarify if this egress flux drop persists.

From the broadband light curve, we derive a transit depth of \HD of $410\pm11$\,ppm in the full 1.1--1.7$\,\si{\micro\meter}$ band observed with HST/WFC3. This results in a planet radius of $2.313\pm0.051\,\si{\rearth}$, assuming the stellar radius given in Table~\ref{tab:ariadne}. This planet radius is slightly larger than the value of $2.16\pm0.08\,\si{\rearth}$ derived by~\citet{Teske2020HD86226c}. While the planet may appear smaller in the wavelength range covered by TESS, the radius difference of less than $2\,\sigma$ does not allow for definite conclusions. For the analysis of the planet spectrum in Sect.~\ref{sec:analysis}, we assumed a planet radius of $2.313\pm0.051\,\si{\rearth}$, as derived from the HST data.

We explored different variations of the treatment of the row shift decorrelation parameter for the fitting of the spectroscopic light curves. We tested whether it is possible to fix the slopes of the row shift correction to the values of the broadband light curve. Except for three bins, the BIC strongly favors leaving these slopes as free parameters for the spectral light curve fits (see Table~\ref{tab:red:BIC}). Furthermore, we tested whether fitting the linear slopes with row shift per visit is useful, analogous to the remaining systematics. As this adds 16 additional parameters to the systematics models, the BIC disfavors this approach even for the broadband light curve. Removing the correction for the row shift in the systematics model increases the BIC in every wavelength bin between 28 and 543, demonstrating that it is necessary to consider this effect when fitting the data.

\begin{figure}[t]
	\centering
	\includegraphics[width=0.49\textwidth]{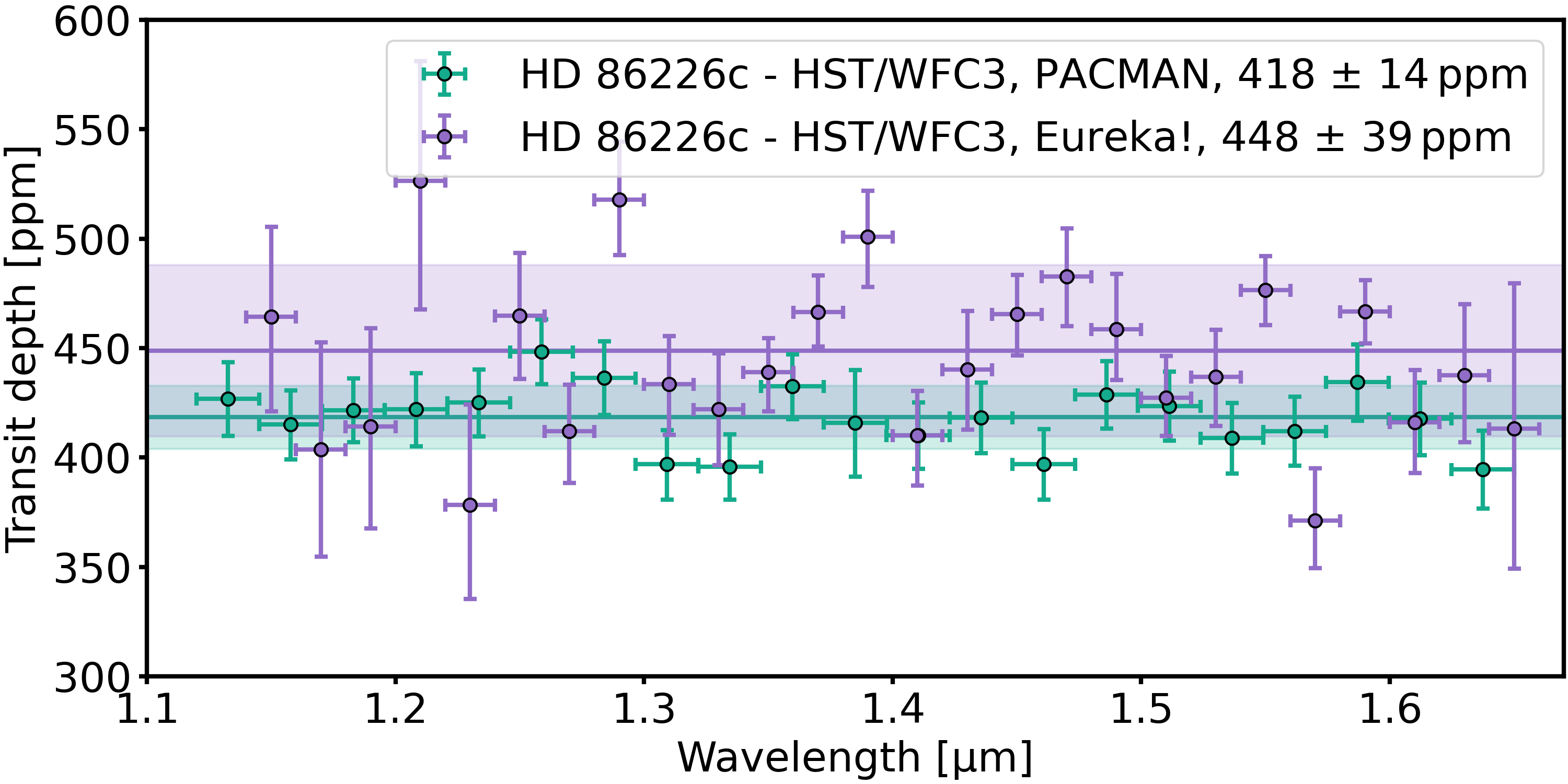}
	\caption{Near-infrared transmission spectrum of \HD obtained with \texttt{PACMAN} (green) and \texttt{Eureka!} (purple). The corresponding shading indicates the $\pm1\,\sigma$ range of a constant transit depth fit to each spectrum.}
	\label{fig:red:spectrum}
\end{figure}

To obtain the spectrum of the planet, we divided the signal into 21 spectral bins between $\SI{1.12}{\micro\meter}$ and $\SI{1.65}{\micro\meter}$. The transit mid-point was fixed to the value derived from fitting the broadband light curve, while transit depth and all systematics were free parameters. Limb darkening coefficients were calculated for each wavelength bin using the \texttt{ExoTiC-LD}~\citep{Grant2024ExoticLD} package for Python. The spectrum is shown in Fig.~\ref{fig:red:spectrum} and Appendix~\ref{app:allan_pacman} lists the transit depths of the individual spectral light curves. We do not see major contributions of red noise to the RMS of the spectroscopic light curve fits with \texttt{PACMAN} (corresponding figures are available via \href{https://zenodo.org/records/16035649}{Zenodo}). Possible atmospheric models that can explain this spectrum are discussed in Section~\ref{sec:models}.

\begin{table}[tb]
    \centering
    \caption{Difference in BIC between discarded and applied systematic models for fitting the light curves with \texttt{PACMAN}.}
    \label{tab:red:BIC}
    \begin{tabular}{crrrr}
         \hline
         Bin & no RS & untied RS & white RS & pol2 \\
         \hline
         Broadband & 751.0 & 11.3 & - & -78.4 \\ 
         1.13 & 96.1 & 105.3 & 73.5 & 35.30 \\
1.16 & 141.9 & 81.8 & 5.8 & 36.70 \\
1.18 & 48.7 & 86.2 & -1.9 & 42.60 \\
1.21 & 30.2 & 94.6 & 0.8 & 41.20 \\
1.23 & 96.4 & 89.9 & -12.4 & 40.40 \\
1.26 & 59.2 & 94.4 & 31.7 & 42.40 \\
1.28 & 129.9 & 88.3 & 10.1 & 37.60 \\
1.31 & 542.7 & 17.0 & 414.2 & -20.40 \\
1.33 & 77.9 & 68.7 & -3.6 & 28.50 \\
1.36 & 220.3 & 75.2 & 90.0 & 23.40 \\
1.39 & 101.7 & 86.4 & 10.4 & 47.60 \\
1.41 & 190.5 & 64.6 & 86.7 & 8.10 \\
1.44 & 402.4 & 51.6 & 286.3 & 16.90 \\
1.46 & 362.7 & 67.8 & 229.8 & 27.40 \\
1.49 & 142.7 & 45.7 & 46.4 & -1.00 \\
1.51 & 28.2 & 72.3 & 9.5 & 24.10 \\
1.54 & 253.1 & 41.7 & 134.6 & -12.10 \\
1.56 & 56.9 & 50.0 & -2.5 & -7.20 \\
1.59 & 202.6 & 28.3 & 139.4 & -18.10 \\
1.61 & 149.3 & 62.1 & 3.8 & 15.20 \\
1.64 & 188.8 & 56.4 & 137.7 & 14.30 \\
         \hline\hline
    \end{tabular}
    \tablefoot{The \texttt{PACMAN} systematics model for the light curve is compared to alternative modeling approaches: without row shift (RS) correction (Col.~2), with an individual RS correction per visit (Col.~3), with RS parameters fixed to the results of the broadband fit (Col.~3), and with a second order polynomial of the flux with time per visit (Col.~4). A negative value suggests that a model is statistically preferred over the applied systematics model for a respective bin.}
\end{table}

\subsection{Light curve fitting with \texttt{Eureka!}}\label{subsec:LCwithEureka}
The data reduced with \texttt{Eureka!} were also simultaneously fit with a \texttt{batman}~\citep{Kreidberg2015batman} transit model and a systematics model that was applied multiplicatively. In contrast to \texttt{PACMAN}, \texttt{Eureka!} was not designed for analyzing multiple transits simultaneously, so we implemented a custom method to perform a joint fit. 

This method shared the fit parameters of the astrophysical model across all visits, while allowing the parameters of the systematics model to be freely determined for each visit. With this setup, each iteration of the fitter determined the best-fit systematics model for each visit individually while keeping a shared astrophysical model. The best-fitting model was determined by the combination of systematic and astrophysical parameters that yielded a minimized deviation between data and model across all visits.

The two scan directions were fit separately to account for the offset between forward and reverse scans. We determined the final transit depth by calculating a weighted average between the fits to the two scan directions. If the flux offset between scan directions were constant, normalizing the respective light curves would lead to identical transit depths. As the spectrum shift on the detector leads to a non-constant offset between both scan directions, this averaging provides an intermediate solution between both directions without correcting for the shift. 

The astrophysical model had the transit midpoint and the transit depth as the only free parameters. The remaining orbital parameters were fixed to the literature values derived by~\citet{Teske2020HD86226c} and~\citet{Kokori2023Exoclock}, analogous to the \texttt{PACMAN} light curve fit. A quadratic limb darkening law with fixed coefficients from the~\citet{Castelli2003_stellarmodel} stellar grid was used. 

The systematics model contained a first-order polynomial per visit and the \texttt{hstramp} model implemented in \texttt{Eureka!}. This model fits an exponential ramp to each orbit according to:
\begin{equation}
    h_0\times \exp(-h_1 \times t_\mathrm{batch}), \quad t_\mathrm{batch} = (t_\mathrm{local}-h_5) \% h_4.
\end{equation}
In this equation, $h_0-h_5$ are free parameters, and $t_\mathrm{local}$ is the time since the visit started. \texttt{Eureka!} furthermore provides the capability to fit a second-order polynomial per orbit with the parameters $h_2-h_3$, which we did not use. An additional parameter, $h_6$, was varied as a fixed parameter with possible values of $0.2$, $0.3$, or $0.4$. It fits the timing offset of the \texttt{hstramp} model. 

A range of other correction techniques were tested, including individually and simultaneously fitted transits, positional adjustments (with the inbuilt functions \texttt{xpos}, \texttt{ypos}, \texttt{ywidth}), and separate consideration of forward and reverse scan directions, where spectra generated by only single scan directions were evaluated. While effective under more stable conditions, these methods proved challenging to implement for this dataset. Isolating data segments by removing integrations or entire transits did not fully resolve the issues.

The spectroscopic light curves were fit using the \texttt{divide\_white} method~\citep[see e.g.,][]{Kreidberg2014GJ1214b}. In addition to fitting the systematics model of the respective wavelength bin, the data were further corrected by the residuals between the data and the broadband light curve model. This was necessary to remove the remaining red noise seen in the broadband light curve (see Appendix~\ref{app:allan_eureka} and the corresponding figures available via \href{https://zenodo.org/records/16035649}{Zenodo}). Furthermore, the transit mid-time was fixed to the value derived from the broadband light curve fit.

Parameter estimates and uncertainties for the fits were obtained using a MCMC algorithm for the light curve fits with the python package \texttt{emcee}~\citep{ForemanMackey2013emcee}. The final light curve fits still show a large contribution of red noise in most of the 27 spectral bins between $\SI{1.12}{\micro\meter}$ and $\SI{1.66}{\micro\meter}$. A major contributor to this systematic noise is the inclusion of the first orbit of every visit. The exponential ramp of this orbit is not identical to the ramp in the subsequent orbits, even after removing different amounts of exposures at the start of this orbit. Unfortunately, removing the first orbit of each visit in this reduction was impossible, as the many half-covered transits terminated the fitting with \texttt{Eureka!} in this case. The transit depth of the first wavelength bin between $\SI{1.12}{\micro\meter}$ and $\SI{1.14}{\micro\meter}$ deviates strongly from the remaining bins, which is why we remove it from further analysis.

Due to the remaining red noise, the uncertainties on the transit depths derived with \texttt{Eureka!} are underestimated. To account for this, we applied a $\beta$ scaling similar to the methods explained by~\citet{Pont2006beta} and~\citet{Cubillos2017beta} to our results. The resulting spectrum with inflated uncertainties is shown in Fig.~\ref{fig:red:spectrum}, and the inflation method is detailed further in Appendix~\ref{app:allan_eureka}.

\subsection{Differences between the spectra obtained with \texttt{PACMAN} and \texttt{Eureka!}}\label{subsec:PACMANandEureka}

The spectra obtained with \texttt{PACMAN} and \texttt{Eureka!} are both featureless but offset by 30\,ppm. In addition, the \texttt{Eureka!} data points show a larger scatter, especially at small wavelengths. Major differences can be found in the light curve fitting processes of both pipelines: We included different data, chose different systematics treatments, and corrected for the trace position with the \texttt{PACMAN} pipeline only.

The offset in transit depth is caused by the inclusion and exclusion of the first orbit of every visit by \texttt{Eureka!} and \texttt{PACMAN}, respectively. The ramp shape of this first orbit differs from the subsequent ramps, which introduces a systematic offset in transit depth when all orbits are treated equally. Including the first orbit of every visit in the \texttt{PACMAN} light curve fits also leads to an increased transit depth comparable to the average value in the \texttt{Eureka!} spectrum.

One cause for the increased scatter in the \texttt{Eureka!} reduction might be the use of the \texttt{divide\_white} method: This method assumes that the fitted systematics model is wavelength independent. However, the spectroscopic fits with \texttt{PACMAN} show that this is not the case. Especially, the first-order polynomials of the baseline flux as a function of time are strongly dependent on the underlying stellar flux that changes continuously with wavelength. 

Another difference between the systematics models is how we account for the upstream-downstream effect and the associated flux dependence on the spectral trace position. While \texttt{PACMAN} jointly fits all data with a constant offset between forward and reverse scans of a visit, \texttt{Eureka!} fits the scan directions separately. Including more data in a single fit might be beneficial for robustly determining the parameters of the complex underlying model. The row shift correction implemented in \texttt{PACMAN} significantly improved the light curve fits and reduced the scatter in the derived spectrum. However, the derived spectroscopic transit depths deviate from the uncorrected values by less than $1\,\sigma$. This correction therefore only partially accounts for the differences in the spectra.

To further investigate the reasons for the deviating spectra, we also inspected the raw light curves. We noticed that these already show an average systematic deviation of 600\,ppm in the normalized flux. While the flux in some orbits only shows minor deviations of a few ppm, other orbits show a systematic flux offset across the full orbit. This is indicative of possible differences in the spectrum extraction. While both pipelines implemented an optimal extraction algorithm, the extraction regions differ. \texttt{Eureka!} uses a fixed, user-assigned region, while \texttt{PACMAN} determines the optimal extraction region individually for each exposure. Given the trace position instability in and between orbits (see Sect.~\ref{subsec:pacmam}), this individual treatment of exposures is important for our data. Furthermore, the background estimations differ: While \texttt{Eureka!} considers pixels outside of a certain row interval, \texttt{PACMAN} considers all pixels in the difference-image of the spectral traces below a threshold value.

These differences in the reduction also lead to different qualities of the light curve fits: While the RMS of the \texttt{PACMAN} spectroscopic light curve fits is typically only about 10\% above the photon noise, the \texttt{Eureka!} RMS is about twice the photon noise in many bins. Based on these findings, we consider the \texttt{PACMAN} results more reliable than those from \texttt{Eureka!}. \texttt{PACMAN} was designed specifically for HST/WFC3, and this is a particularly challenging dataset. The main problems in the \texttt{Eureka!} analysis arose from the bright host star and corresponding poor trace position stability and from the partial transit observations. While these effects were accounted for with \texttt{PACMAN}, \texttt{Eureka!} does not currently provide the required capabilities. Given the featureless nature of the spectrum, implementing these capabilities to \texttt{Eureka!} is out of the scope of this analysis. For further analysis and modeling of the planetary atmosphere, we proceeded with the spectrum obtained with the \texttt{PACMAN} pipeline.
\begin{figure*}[t]
	\centering
	\includegraphics[width=0.59\textwidth]{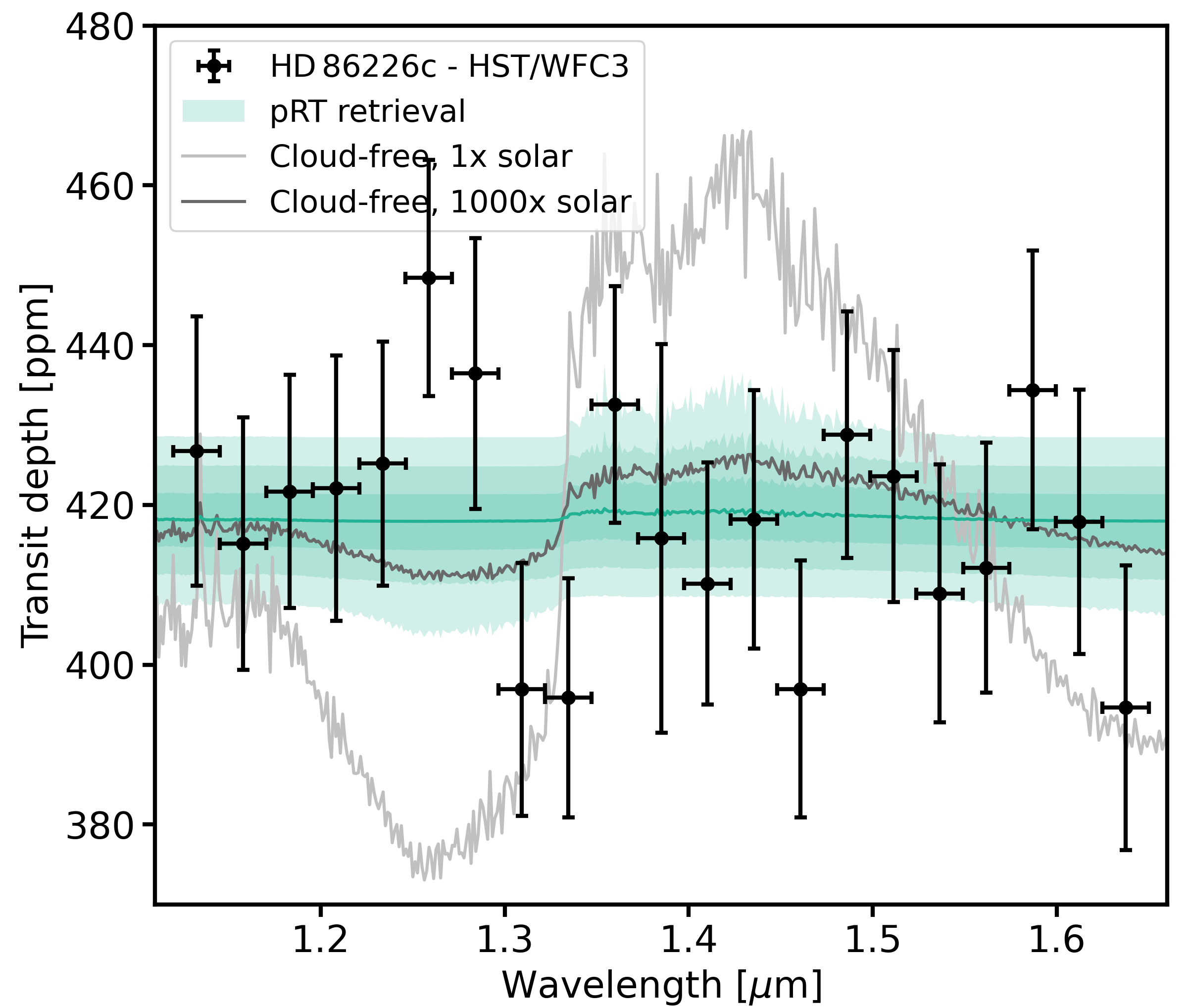}
  \includegraphics[width=0.405\textwidth]{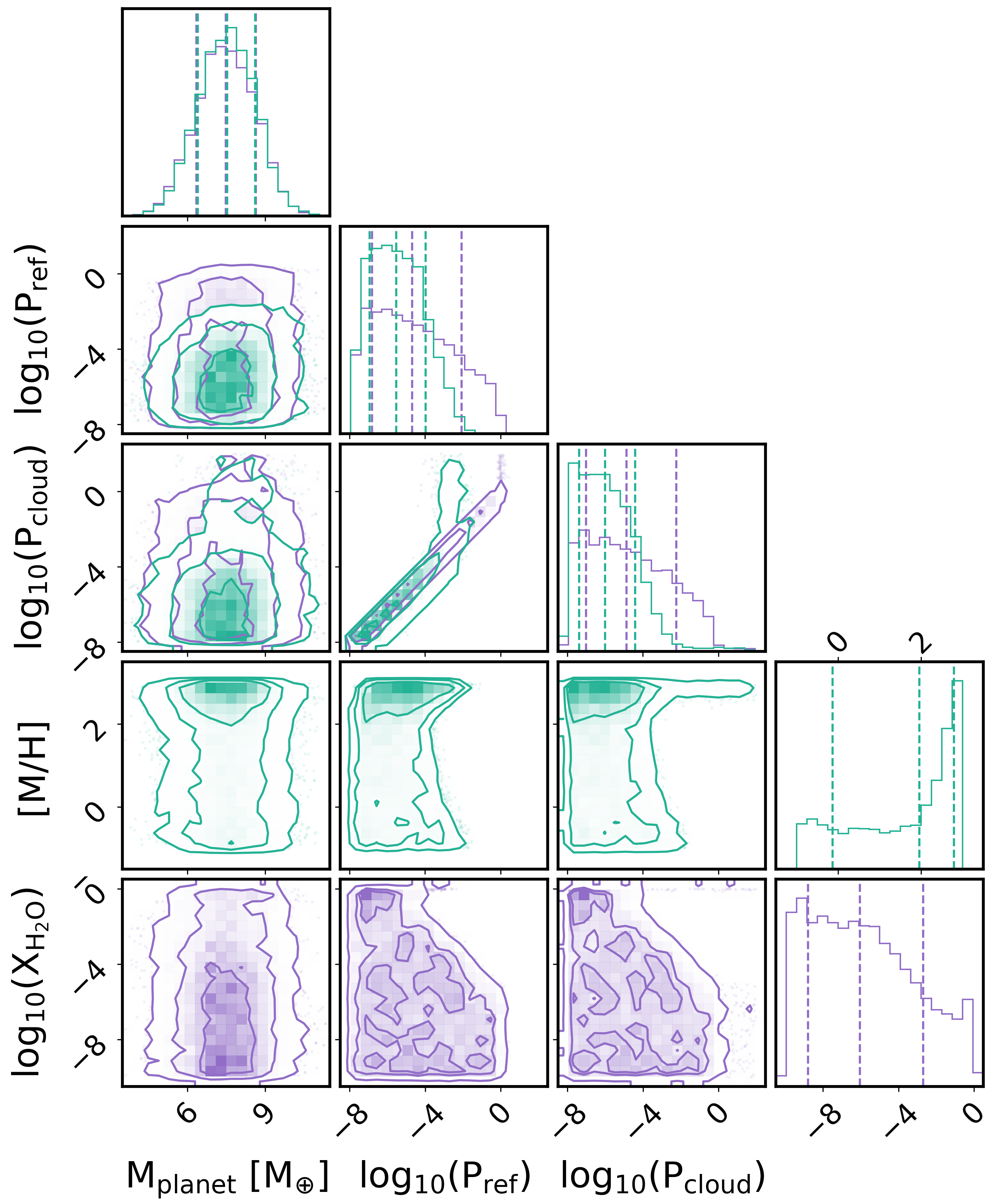}\\
	\caption{Results of the pRT retrievals for the atmosphere of \HD. Left: \texttt{PACMAN} spectrum (black data) and 1, 2, and 3\,$\sigma$ percentiles of the spectra from the posteriors of the pRT retrieval with scaled solar chemistry (green shadings). The gray spectra show cloud-free models with 1$\times$ and 1000$\times$ solar metallicity. Right: Posterior distribution for the two retrieval runs with scaled solar chemistry (green) and H-He-H$_2$O atmosphere (purple). The pressures are provided in units of $\log_{10}([\mathrm{bar}])$. Contours are drawn at the 1, 2, and 3\,$\sigma$ levels. Vertical dashed lines mark the 16\%, 50\%, and 84\% quantiles.}
	\label{fig:prt:output}
\end{figure*}

\section{Transit spectrum models}\label{sec:models}
Given the precision of our measurements, a visual inspection of the transmission spectrum of \HD (see Fig.~\ref{fig:red:spectrum}) indicates that it is likely featureless. As a simple test for the presence of spectral features, we performed a least-squares fit to a constant transit depth with the \texttt{curve\_fit} function of the Python package \texttt{scipy.optimize}~\citep{Virtanen2020SciPy-NMeth,Vugrin2007curvefit}. For the  \texttt{PACMAN} spectrum, this leads to a value of $418\pm14\,\si{\ppm}$, with a $\chi^2$ deviation of $16.6$. According to chi-squared distributions with 20 degrees of freedom, the data is compatible with a constant transit depth within $0.4\,\sigma$. 

The featureless spectrum strongly rules out a clear solar-metallicity atmosphere for \HD, but other scenarios such as high atmospheric metallicity or high altitude clouds are still possible \citep{Benneke2013cloudvsweight}. To quantify these potential scenarios, we used the retrieval package petitRADTRANS~\citep[pRT;][]{Molliere2019pRT,Nadeskin2024pRT} to fit atmospheric models to the spectrum of \HD (see Sect.~\ref{subsec:pRT}). In addition, we ran self-consistent cloud and haze models for the planet, which are described in Sections~\ref{subsec:foward_models_cloud} and~\ref{subsec:foward_models_haze}. 

\subsection{Retrievals with petitRADTRANS}\label{subsec:pRT}
Retrieving a featureless spectrum does not lead to a precise constraint of physical or chemical properties of the atmosphere but rather to a broad range of conditions consistent with the observed spectrum. Due to this, we employed a simplified approach with a minimal number of retrieved parameters. 

To simulate the spectra of the planet, we assumed an isothermal atmosphere with the equilibrium temperature $T_\mathrm{eq}=\SI{1310}{\kelvin}$~\citep{Teske2020HD86226c}. This approximation is sufficient, as we observe the planet in transmission and are only sensitive to a narrow layer of the atmosphere~\citep[e.g.,][]{Fortney20103D}. We simulated the atmosphere in a pressure range between $\log_{10}(P [\mathrm{bar}])=2$ and $\log_{10}(P [\mathrm{bar}])=-8$. To calculate the transmission spectrum, pRT first converts the pressure structure of the atmosphere to a radius structure, assuming hydrostatic equilibrium. To specify this conversion, it is necessary to define a pair of reference pressure, $P_\mathrm{ref}$, and reference radius, $R_\mathrm{ref}$. Based on the transit depth of the broadband light curve (see Sect.~\ref{subsec:LCwithPACMAn}), we fixed the reference radius to $R_\mathrm{ref}=\SI{2.313}{\rearth}$. The reference pressure was left as a free parameter with a log-linear prior across the entire pressure range to account for offsets in the transit depth. We retrieved the planet mass with a Gaussian prior centered on the value $\SI{7.25}{\mearth}$ with the $1\,\sigma$ width of $\SI{1.19}{\mearth}$, following the value obtained by~\citet{Teske2020HD86226c}. Lastly, to calculate the transit depths, we fixed the stellar radius to $R_\mathrm{star}=\SI{1.047}{\rsun}$, as derived from the \texttt{ARIADNE} models (see Sect.~\ref{subsec:ariadne}).

We also included a gray cloud with variable pressure in the retrieval. This was implemented such that the atmosphere is fully transparent at altitudes above the cloud level and fully opaque below. The clouds were placed at the layer with the pressure $P_\mathrm{cloud}$, which was retrieved with a uniform prior across the simulated pressure range between $\log_{10}(P_\mathrm{cloud} [\mathrm{bar}])=2$ and $\log_{10}(P_\mathrm{cloud} [\mathrm{bar}])=-8$. With this cloud implementation, the spectra do not contain absorption signals that would be imprinted at altitudes below the cloud deck. The amplitude of the corresponding absorption features can therefore be muted. 

We used two approaches for the atmospheric chemistry. In the first approach, we assumed that the atmosphere has a scaled solar composition in chemical equilibrium. The abundances were obtained from the pRT subpackage \texttt{chemistry.pre\_calculated\_chemistry}, whose tables were calculated with the \texttt{easyCHEM} python package~\citep{leimolliere2024}\footnote{\url{https://easychem.readthedocs.io/en/latest/}}. Since the spectrum does not inform us about the abundances of individual species, we fixed the C/O ratio to $0.55$. This value was estimated based on the C/H and O/H abundances of HD\,86226 stated in the Hypatia catalogue~\citep{Hinkel2014hypatia} and is equal to the solar value~\citep{Asplund2009sun}. We retrieved the metallicity of the atmosphere 
\begin{equation}
    \left[\frac{\mathrm{M}}{\mathrm{H}}\right]=\log_{10}\left(\frac{N_\mathrm{M}}{N_\mathrm{H}}\right)_\mathrm{planet} - \log_{10}\left(\frac{N_\mathrm{M}}{N_\mathrm{H}}\right)_\odot,
\end{equation}
where $N_\mathrm{H}$ and $N_\mathrm{M}$ are the number fractions of hydrogen and species heavier than helium, respectively. We let [M/H] vary from $-1$ to $3$ and included opacities from H$_2$O~\citep{Polyansky2018exomol}, CO~\citep{Rothman2010hitemp}, CO$_2$~\citep{Yurchenko2020exomol}, and CH$_4$~\citep{Hargreaves2020CH4}.

The second approach assumed a simplified atmosphere consisting only of H$_2$, He, and H$_2$O. We focused on H$_2$O because it has a strong feature in the observed wavelength range between $1.1$ and $\SI{1.7}{\micro\meter}$ and is a species with high expected abundances based on planet formation models~\citep{Fortney2013framework,Bitsch2021watercontent,Izidoro2021pebblemigration,Burn2024valley}. For a planet with such a high equilibrium temperature ($T_\mathrm{eq}=\SI{1310}{\kelvin}$), we do not expect high abundances of CH$_4$ as it is photodissociated~\citep{Zahnle2009soot,Gao2020CH4}. Other species that could be expected in the atmosphere of \HD are CO and CO$_2$~\citep[e.g.,][]{Moses2013composition}. However, these only have shallow features in the observed wavelength range, which is why we neglected them for this analysis. During the retrieval, we fitted for the H$_2$O mass fraction, $X_{\mathrm{H}_2\mathrm{O}}$, and calculated the H and He mass fractions according to their solar ratio~\citep[see][]{Asplund2009sun}. We note that this ratio could differ for sub-Neptunes, as a preferential mass-loss of H over He could alter the atmospheric composition~\citep{Malsky2023helium}. However, our data is not sensitive enough to determine a change in the H-He ratio.

In addition to molecular absorption, we also included collision-induced absorption of H$_2$ and He, as well as Rayleigh scattering from H$_2$, He, and H$_2$O as additional opacity in both approaches. We used the opacities for collision induced absorption between H$_2$ molecules from~\citet{Borysow2001H2H2,Borysow2002H2H2}, and between H$_2$ and He from~\citet{Borysow1988H2He,Borysow1989H2He,Borysoe1989H2HeII}.
Opacities from Rayleigh scattering were computed internally by pRT, based on the Rayleigh scattering cross-sections of H$_2$~\citep{Dalgarno1962H2}, He~\citep{Chan1965He}, and H$_2$O~\citep{Harvey1998H2O}.

The best-fit models and the posterior distribution of the parameters for both chemistry treatments are shown in Fig.~\ref{fig:prt:output}. From these, it is clear that we cannot distinguish whether high-altitude clouds or a high-metallicity atmosphere cause the featureless spectrum. High-altitude cloud models allow for most values of [M/H] since the cloud always hides the corresponding molecular features. Also, models with the highest [M/H] allow the opaque cloud deck to be placed at most pressure levels. For these, the spectrum is sufficiently flat due to the high mean molecular weight alone. At intermediate cloud pressures ($\log_{10}(P_\mathrm{cloud}[\mathrm{bar}])=-3$ to $0$) and metallicities below [M/H]=2, feasible models cluster in a triangle shape, allowing for higher cloud pressures at lower metallicities. This is reasonable, as the respective molecules form absorption features deeper in the atmosphere for low metallicities. Therefore, low-altitude clouds can also sufficiently mute their spectral features.

The reference pressure and gray cloud pressure level have a nearly one-to-one correlation. Both reference and cloud pressure control the effective planet radius in the model and, consequently, the transit depth of the featureless spectrum. The parameters are therefore expected to have similar values. A deviation from this correlation is seen for high cloud pressures. For these, the transit depth is not set by the cloud but mainly by the assumed reference pressure.

Attempts to run the retrievals with a free atmospheric temperature were unreliable, as the models always ran into the lower prior boundaries of the temperature interval. Expanding the prior interval to between $\SI{50}{\kelvin}$ and $\SI{3000}{\kelvin}$ led to a preferred value of $T=\SI{200}{\kelvin}$, much lower than the equilibrium temperature of the planet. These solutions fit the spectrum equally well as the high-temperature models with high metallicity or high-altitude clouds. An atmosphere with a lower temperature also has a lower scale height, decreasing the feature size. For the retrieval, this leads to a large prior volume with equally good fits of low-temperature models, which is then also seen in the posterior. To avoid biasing our results with unphysically low temperatures, we keep the planet temperature in the retrieval fixed at \SI{1310}{\kelvin}.

We also tested our sensitivity to cloud-free atmospheres. To determine the metal enrichment that could produce a sufficiently flat spectrum, we ran the retrievals with scaled solar chemistry again without the gray cloud deck. These retrievals show that the featureless spectrum can also be reproduced with a metallicity of [M/H]=$2.9^{+0.1}_{-0.2}$ and a reference pressure of $\log_{10}(P_\mathrm{ref}[\mathrm{bar}])=-2.6^{+0.5}_{-0.4}$. In the absence of clouds, we derive a $3\,\sigma$ confidence lower limit on the atmospheric metallicity of \HD of [M/H]=$2.3$, corresponding to a minimum mean molecular weight of approximately $6$\,u. We note that the pre-calculated grids of pRT only cover atmospheric metallicities of up to [M/H]=$3$, which is also the upper boundary of our prior range. Therefore, the actual metallicity of the planet may be much higher than the derived lower limit.

To further quantify the significance with which we can rule out a cloud-free solar-metallicity atmosphere, we retrieved the spectrum a final time without the gray cloud deck and a fixed solar [M/H]=0. To avoid artificially muting the feature size by an overly increased planet mass, we furthermore fixed the planet mass to $\SI{7.25}{\mearth}$~\citep{Teske2020HD86226c}. The best-fitting spectrum with a $\chi^2$ of $89.9$ has a reference pressure of $\log_{10}(P_\mathrm{ref})=-1.50^{+0.07}_{-0.07}$. As shown in Fig.~\ref{fig:prt:output}, the cloud-free solar-metallicity model does not fit the observed spectrum well. Based on the derived $\chi^2$ for one free parameter, this model is ruled out with a confidence of $6.5\,\sigma$. We note that including mass as a free parameter only excludes the cloud-free solar-metallicity model with confidence of $2.9\,\sigma$. However, the corresponding retrieval finds a planetary mass of $11.1\pm0.8\,\si{\mearth}$, much higher than the value of $7.25^{+1.19}_{-1.12}\,\si{\mearth}$ found by~\citet{Teske2020HD86226c}.

\begin{figure}[t!]
    \centering
    \vspace{-5pt}
    \includegraphics[width=0.99\linewidth]{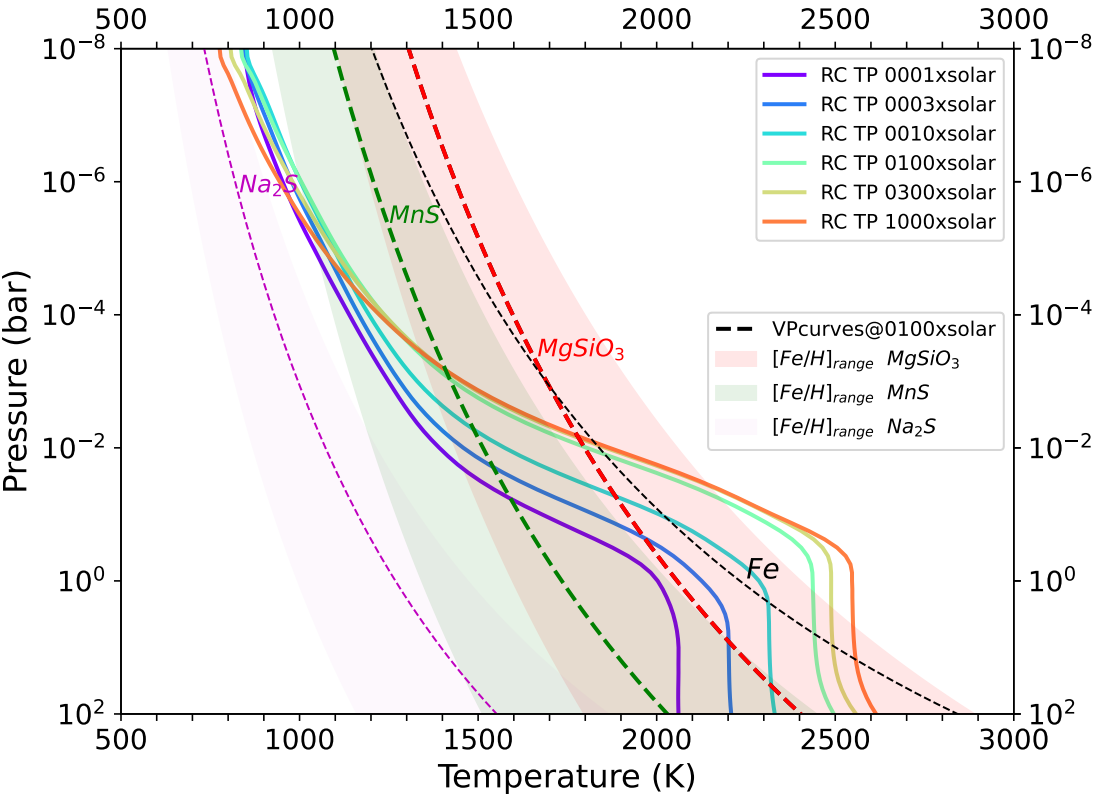}
    \caption{ Radiative-convective temperature-pressure profiles and vapor-pressure curves. Solid lines with rainbow colors show radiative-convective temperature profiles for metallicities from 1 to 1000$\times$solar. This figure also shows the vapor-pressure curves of the species we considered in our analysis that could condense on the temperature regimes of HD\,86226\,c. The vapor-pressure curves at 100$\times$ solar atmospheric metallicity are given as dashed lines, while the extent of the vapor-pressure curves over the range of 1 to 1000$\times$ solar for MnS, MgSiO$_3$, and Na$_{2}$S are given as shaded regions. The Fe range is not shown to avoid clutter, as it largely overlaps the MgSiO$_3$ range. Na$_{2}$S vapor-pressure curves do not cross our temperature-pressure profiles for any atmospheric metallicity. Thus, based on our thermal-stability approach, these clouds could not condense in the atmosphere of HD\,86226\,c and we did not include them in our analysis.}
    \label{fig:JB_VP}
\end{figure}

\begin{figure*}[t!]
    \centering
    \vspace{-5pt}
    \includegraphics[width=0.3\linewidth]{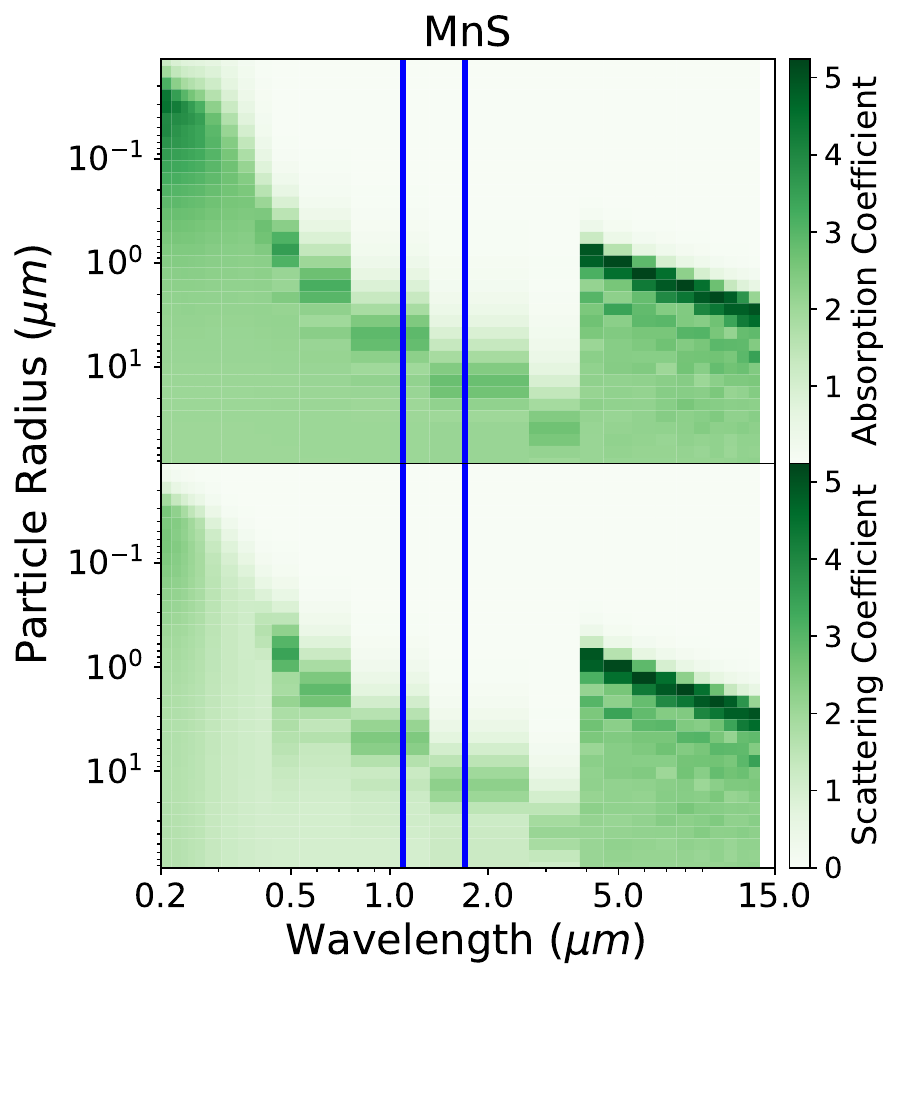}
    \includegraphics[width=0.3\linewidth]{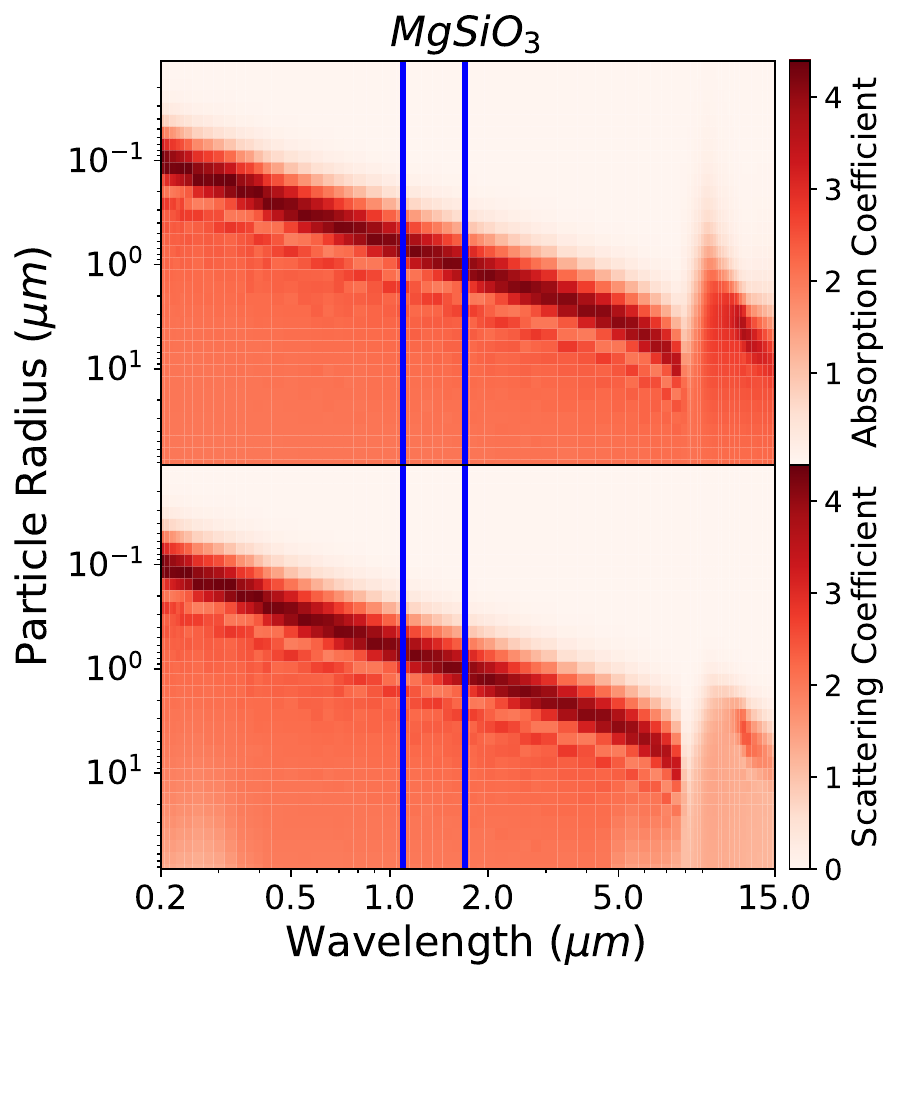}
    \includegraphics[width=0.3\linewidth]{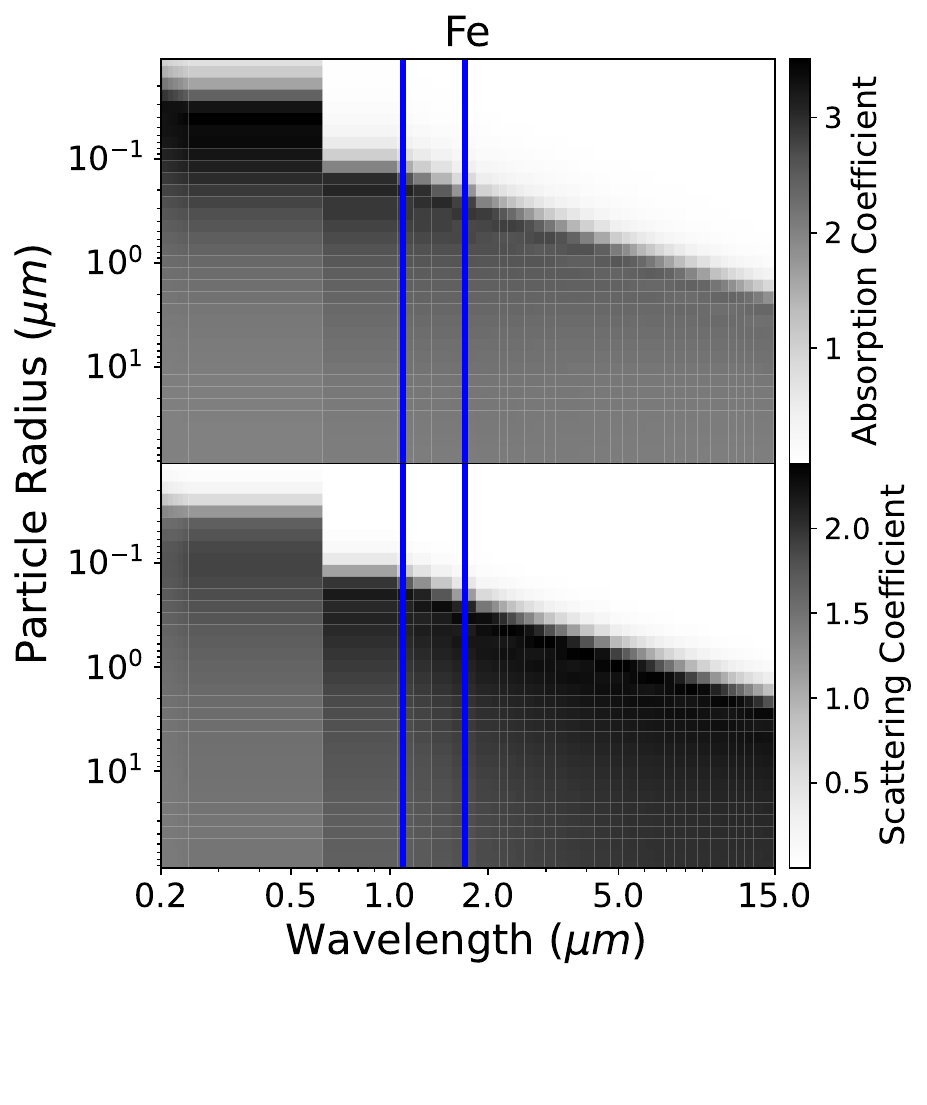}
    \caption{Cloud scattering and absorption coefficients of MnS, MgSiO$_3$, and Fe condensates. All species are featureless on the narrow wavelength range of our observations, 1.1 -- 1.7\,$\si{\micro\meter}$ (marked with blue vertical lines), for the expected sizes of the cloud droplets of 0.01 -- 100\,$\si{\micro\meter}$. As shown, this is not the case for MnS and MgSiO$_3$ clouds in the range of 0.2 -- 15\,$\si{\micro\meter}$, where their effect on the spectra should be distinguishable with observations that have sufficiently high resolution. We note that values are plotted in log-log space to highlight the narrow wavelength range of our observations.}
    \label{fig:JB_Mieff_coeffs}
\end{figure*}

\subsection{Cloud forward models}\label{subsec:foward_models_cloud}
We also explored cloud forward models of metal-enriched atmospheres to investigate plausible cloud compositions. To do this, we used the Thermal Stability Cloud (TSC) model~\citep{Blecic2025TSC}. This model was already successfully applied to multiple synthetic and actual HST and JWST targets to distinguish various cloud species present in their atmospheres and constrain the overall cloud structure \citep[e.g., ][]{venot_global_2020, TaylorEtal_2023, BellEtal_WASP43b_2024}. TSC is a complex Mie-scattering cloud model that is built on the approaches of \cite{Benneke2015arxivSCARLETretrieval} and \cite{AckermanMarley2001apj}, allowing for additional flexibility in the location of the cloud base (determined by the overall atmospheric metallicity) and in the width of the log-normal distribution (which affects the droplet sizes and distribution). The model formulation employs multiple free parameters, enabling us to capture the complex structure of the clouds. These include the cloud's shape, location, and extent; the droplet sizes, distribution, and volume mixing ratio; and the overall atmospheric metallicity. This parametrized model assumes thermally stable equilibrium atmospheric conditions \citep{marley2013clouds} and does not account for non-equilibrium microphysical processes. For more details on the model parameters, refer to the model description in \citet{venot_global_2020}, as well as \citet{AckermanMarley2001apj} and \citet{Benneke2015arxivSCARLETretrieval}.

Before applying the TSC cloud model, we first generated 1D self-consistent cloud-free temperature profiles over a range of different metallicities, [M/H] = 0.0, 0.5, 1.0, 1.5, 2.0, 2.5, and 3.0 (1, 3, 10, 30, 100, 300, and 1000$\times$ solar), applying both radiative-convective and chemical equilibrium calculations (Cubillos, Blecic et al., in prep). Our radiative equilibrium approach followed the iterative procedure of \citet{MalikEtal2017ajHELIOS} and \citet{HengEtal2014apjsTwoStreamRT} and utilized the system parameters provided in Table~\ref{tab:ariadne} and the stellar spectrum derived in Section~\ref{subsec:star_reduction} as input. The approach assumed the presence of all major molecular species formed from these most abundant elements: H, He, C, O, N, Na, K, S, Si, Fe, Ti, and V; and the following ions: e$^-$, H$^-$, H$^+$, H$_2^+$, He$^+$, Na$^-$, Na$^+$, K$^-$, K$^+$, Fe$^+$, Ti$^+$, V$^+$. We set the planetary interior temperature of HD\,86226\,c to $T_\mathrm{int}$ = 100\,K and the energy redistribution factor to $f$ = 2/3~\citep{Hansen2008redist}. The model included the opacities from alkali metals, Na and K, and molecular species H$_2$O, CH$_4$, NH$_3$, HCN, CO, CO$_2$, C$_2$H$_2$, OH, SO$_2$, H$_2$S, and SiO \citep[see][for all relevant references and the line-lists sources]{CubillosBlecic-PyratBay}. Fig.~\ref{fig:JB_VP} shows these cloud-free temperature profiles color-coded by metallicity.

After obtaining the self-consistent temperature profiles for all atmospheric metallicities, we investigated which cloud species could condense within the \HD temperature regimes. These include the species KCl, ZnS, Na$_2$S, MnS, SiO$_2$, Cr, MgSiO$_3$, Fe, and Mg$_2$SiO$_4$~\citep[e.g.,][]{gao2021aerosols, MorleyEtal2012}. We excluded KCl, ZnS, Na$_2$S, Mg$_2$SiO$_4$, Cr, and SiO$_2$ from our analysis for the following reasons: KCl, ZnS, and Na$_2$S condense at lower temperatures and do not intersect the temperature-pressure ($T-P$) profiles for the metallicity range we considered, preventing cloud formation based on our thermal-stability approach. Mg$_2$SiO$_4$ condenses at slightly higher temperatures than MgSiO$_3$. At the temperatures of HD\,86226\,c, MgSiO$_3$ condensates form first and be more abundant~\citep{lodders2006chemistry}. We note that Cr is much less abundant than MnS or MgSiO$_3$, thus significantly reducing the likelihood of Cr cloud formation \citep{Lodders2003, MorleyEtal2012}. In the presence of both Mg and Si elemental species, \cite{VisscherEtal2010} suggest that MgSiO$_3$ forms before SiO$_2$, making SiO$_2$ less likely to form at all. Additionally, neither Cr nor SiO$_2$ has prominent features at the wavelengths of our observations~\citep{Blecic2025TSC}, making them indistinguishable from other, more likely, cloud species. 

Figure~\ref{fig:JB_VP} shows the vapor-pressure ($V-P$) curves of the species we included in our analysis, which could condense between temperatures of 1000–3000 K and pressures of 100–10$^{-8}$\,bar. It is apparent that the position of the $V-P$ curves in the $T-P$ space strongly depends on the atmospheric metallicity \citep[e.g.,][]{MorleyEtal2012} shifting to lower temperatures for lower solar metallicity values and to higher temperatures for super-solar metallicities. We show this range as shaded regions for several cloud species. The plot demonstrates that the MnS, MgSiO$_3$, and Fe $V-P$ curves intersect with the $T-P$ curves for our explored metallicity and pressure range. Even at a metallicity of 1000$\times$ solar, Na$_{2}$S clouds could condense only at extremely low pressures, where their abundance is too low to produce visible signatures in the planetary spectra. Thus, we only proceeded with our grid exploration for the MnS, MgSiO$_3$, and Fe clouds. In Fig.~\ref{fig:JB_Mieff_coeffs}, we also present the scattering and absorption coefficients for these species. While MnS and MgSiO$_3$ exhibit distinct features within the displayed wavelength range, they, along with Fe, are nearly indistinguishable in the range of our observations (1.1 -- 1.7 $\mu$m). This lack of differentiation has posed a challenge for our analysis, as discussed below.

\begin{figure*}[tb]
    \centering
    \includegraphics[width=0.45\linewidth, trim = 4mm 4mm 3mm 8mm, clip]{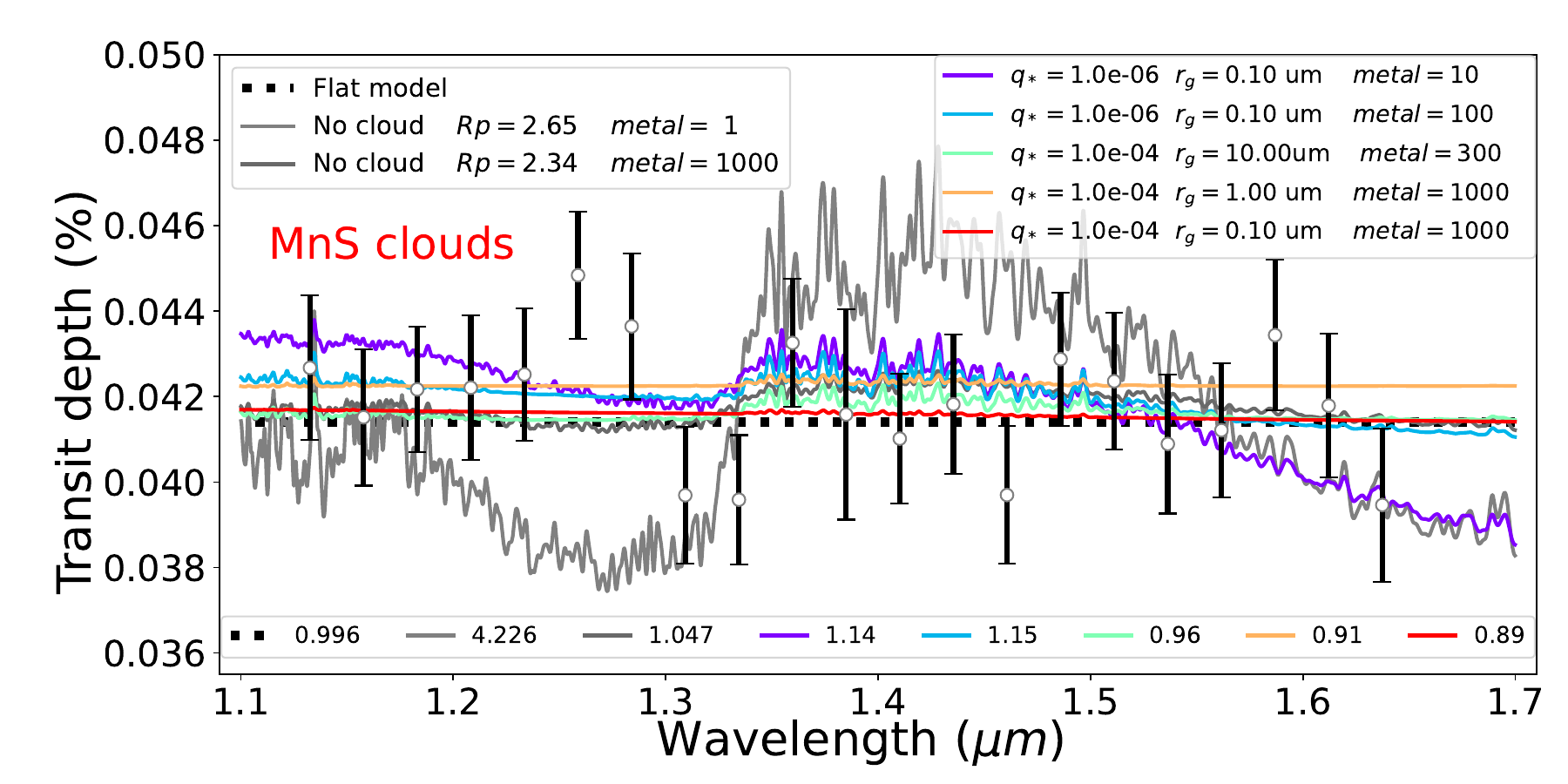}\hspace{-3pt}
    \includegraphics[width=0.55\linewidth, trim = 5mm 1mm 0mm 0mm, clip]{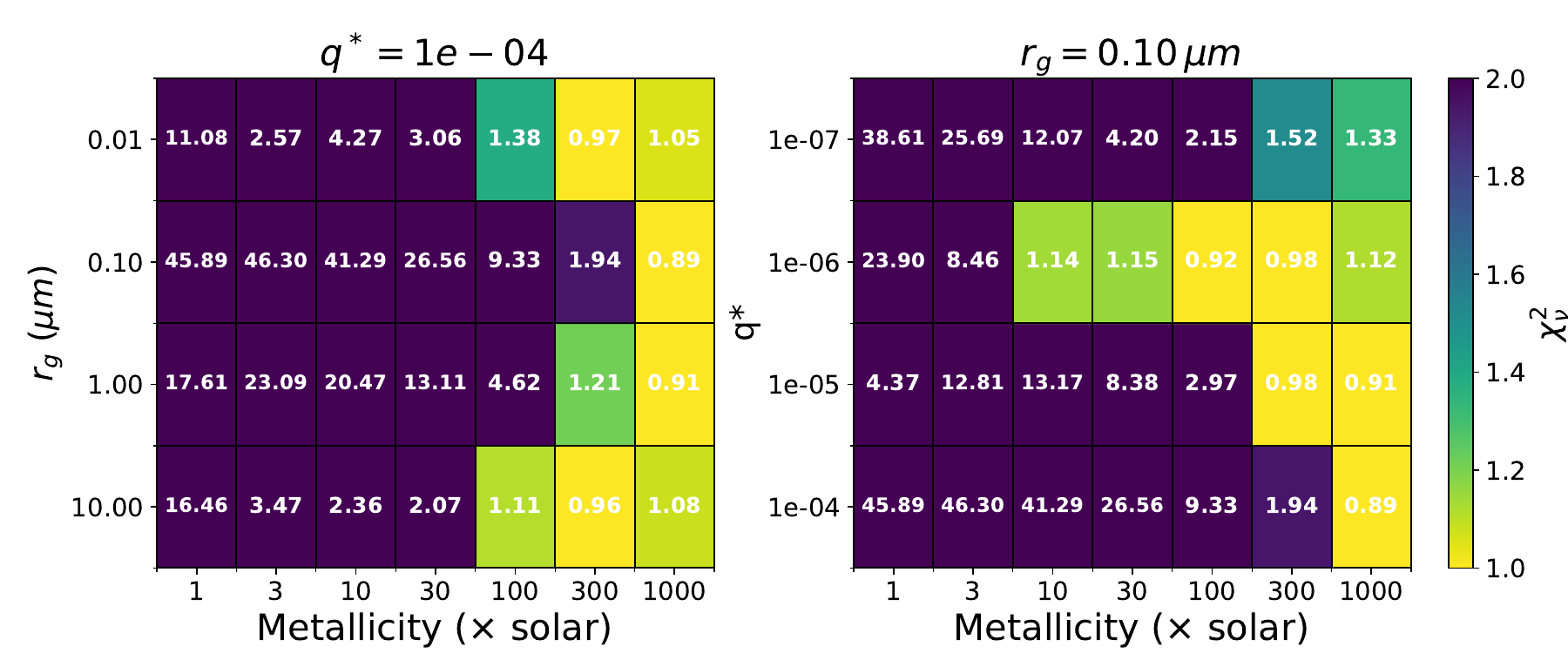}
    \includegraphics[width=0.45\linewidth, trim = 4mm 4mm 3mm 8mm, clip]{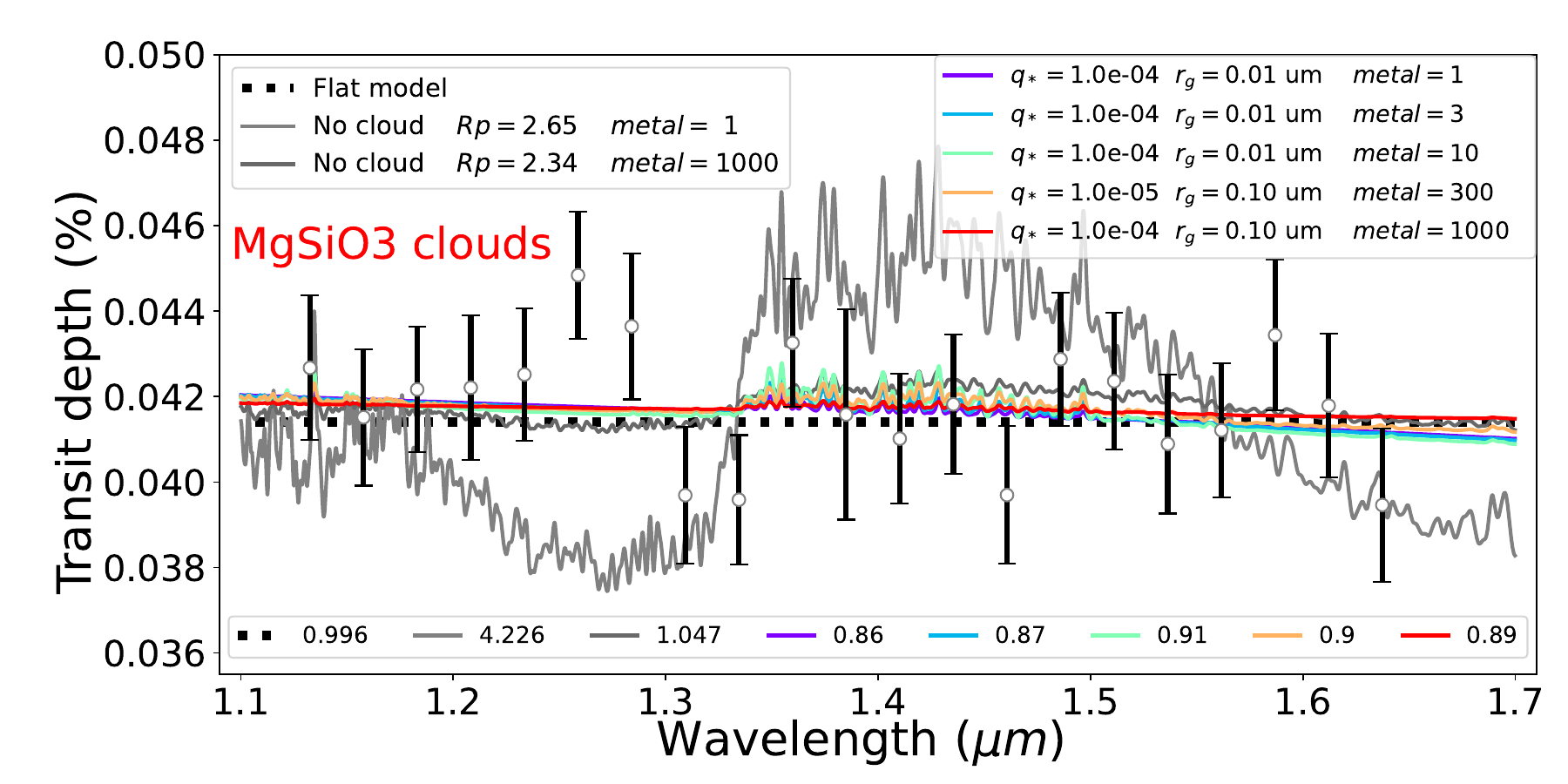}\hspace{-3pt}
    \includegraphics[width=0.55\linewidth, trim = 5mm 1mm 0mm 0mm, clip]{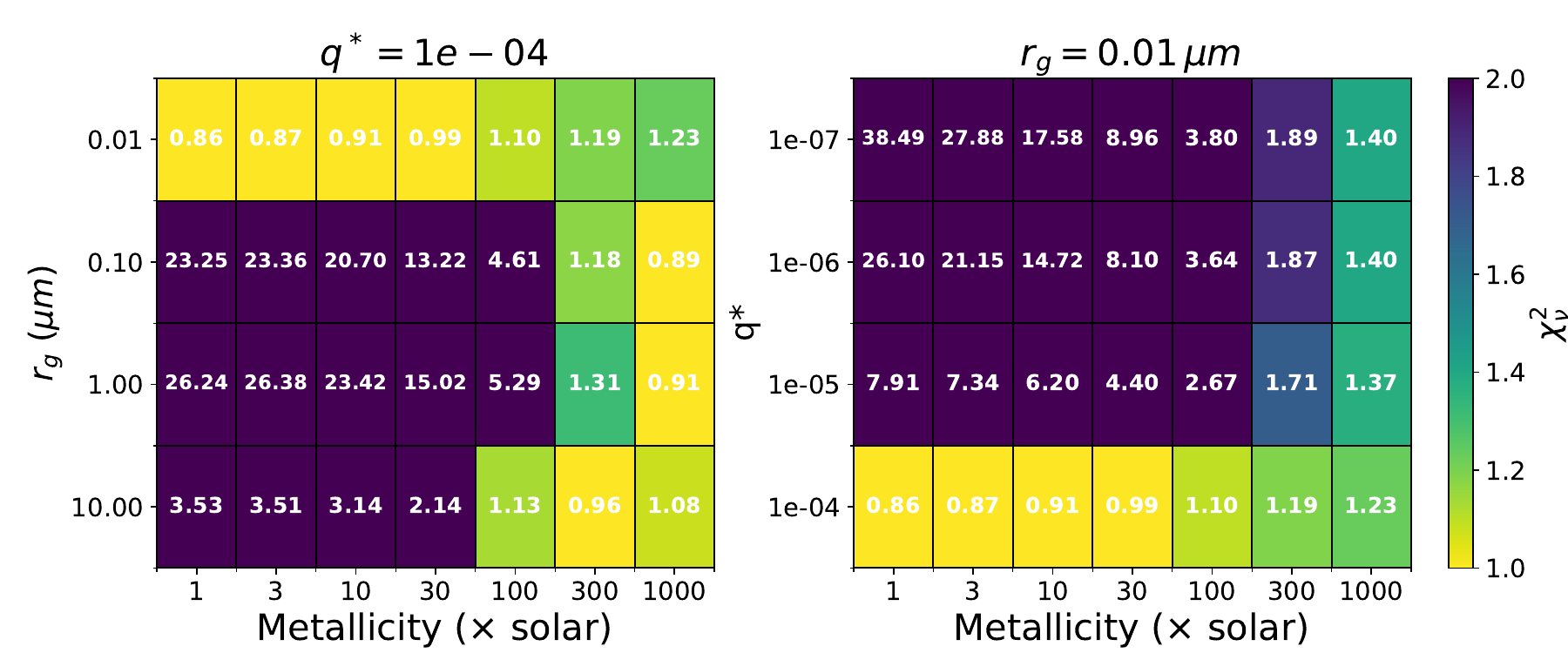}
    \includegraphics[width=0.45\linewidth, trim = 4mm 4mm 3mm 8mm, clip]{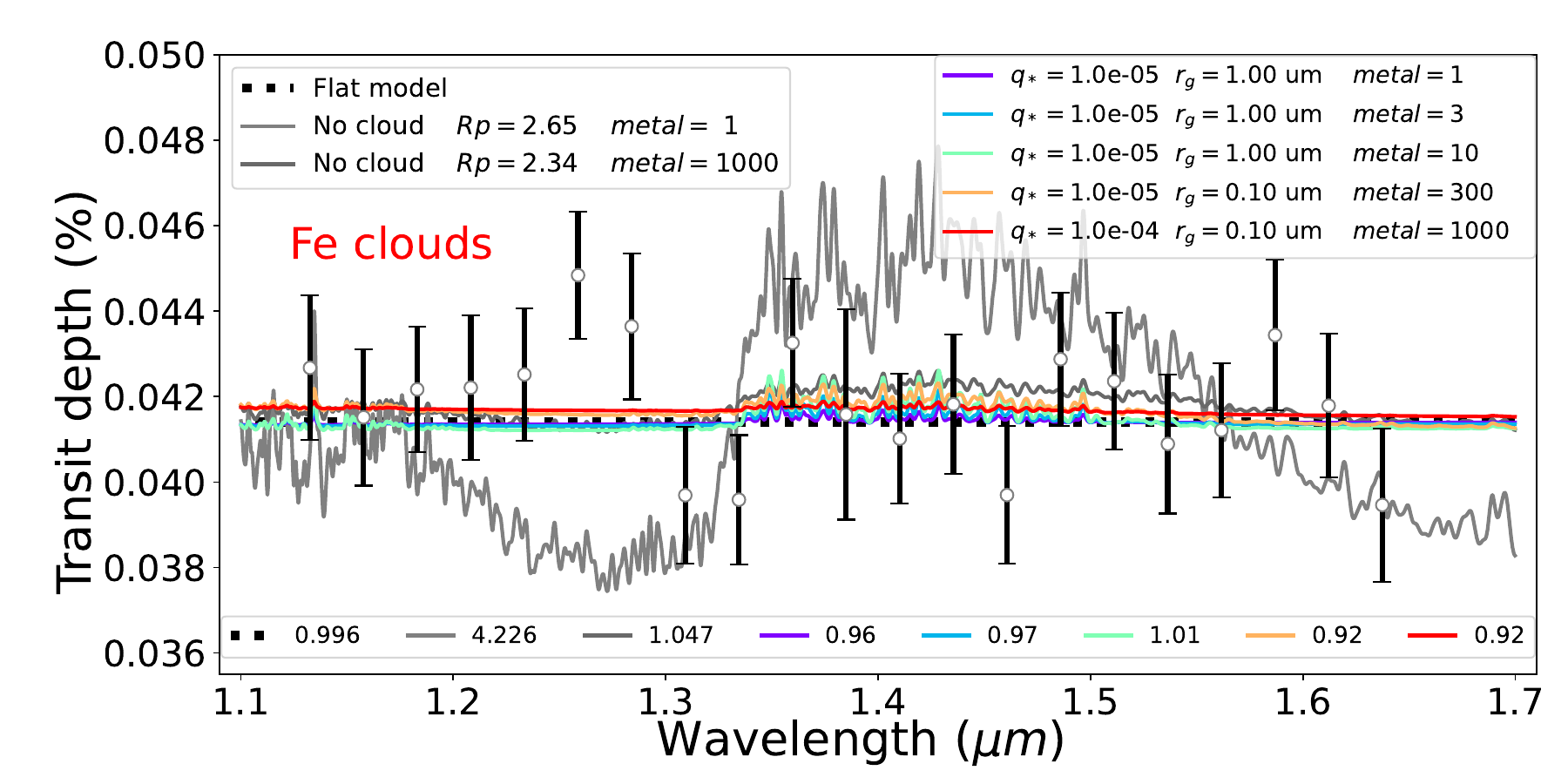}\hspace{-3pt}
    \includegraphics[width=0.55\linewidth, trim = 5mm 1mm 0mm 0mm, clip]{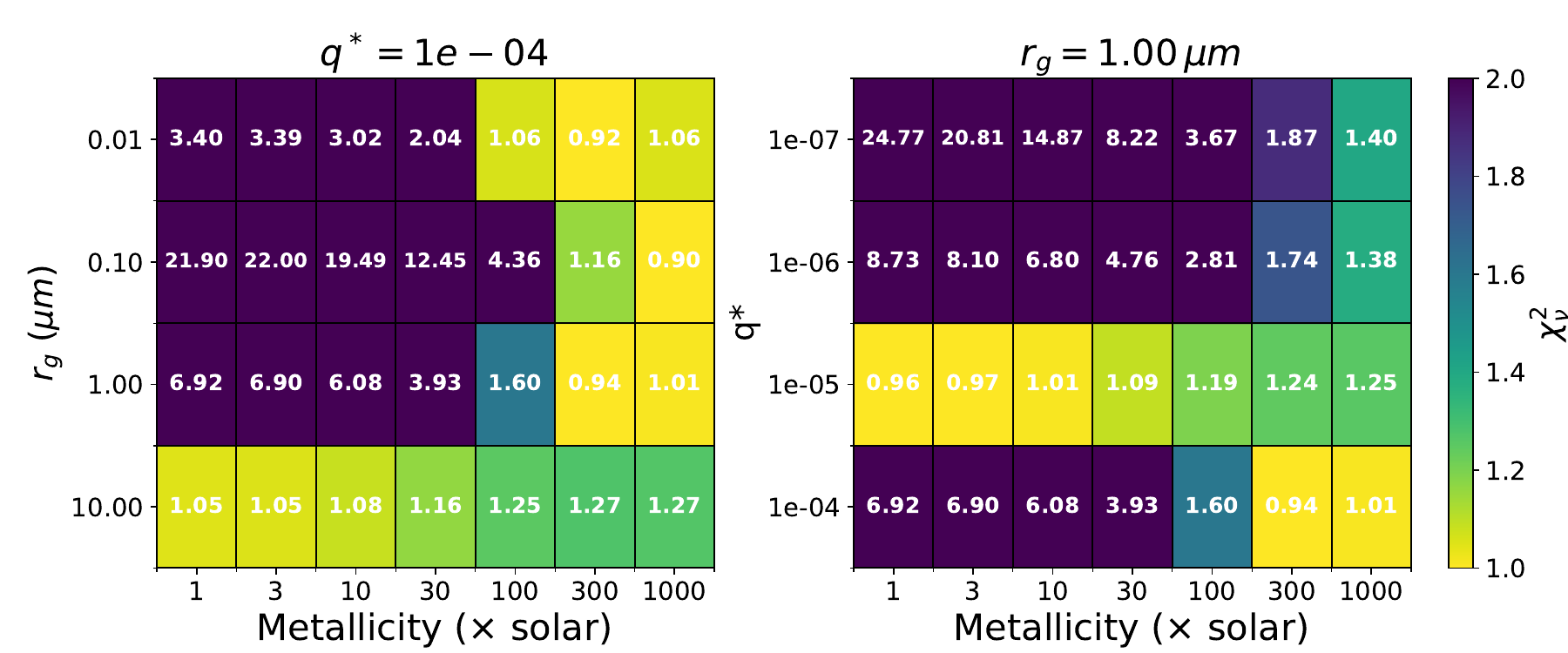}
    \caption{ Comparison between different grid models and the goodness of fit for the MnS (top row), MgSiO$_3$ (middle row), and Fe (bottom row) clouds calculated at the planetary radius set to 2.313\,$\si{\rearth}$. For each species, the left panel shows our observations with uncertainties compared against some selected lowest reduced chi-squared ($\chi^2_\nu$) models, with the corresponding $\chi^2_\nu$ values in the bottom legend. The parameter values for clear, no-cloud models are given in the left legend, and for cloudy models, they are given in the right legend. The flat model is drawn as a dotted line at a transit depth of 418\,ppm. The tables in the middle and right panels display the models' $\chi^2_\nu$ values on a grid of cloud droplet radius $r_g$, droplet volume mixing ratio $q^*$, and metallicity [M/H]. The table titles list fixed parameters. The color bar extent is set between 1 and 2 to highlight only the models that best fit the data. }
    \label{fig:JB_MnS_MgSiO3_Fe}
\end{figure*}

We generated a grid of cloud models across all metallicities, varying fundamental cloud parameters that can produce visible signatures in the spectra. In addition to metallicity, these include droplet radius, $r_g$, and droplet volume mixing ratio, $q^*$, described by the number of cloud droplets over the number of H$_2$ molecules per unit volume. The presence of clouds effectively increases the radius of the planet and decreases the atmospheric scale height. A similar effect on the spectra could result from uncertainties in the planetary transit radius, potentially leading to either an underestimation or overestimation of the cloud droplet volume mixing ratio and droplet size. For this reason, we made the transit radius a free parameter in our model, and we also allowed for a vertical offset of the observed spectrum.

We only considered plausible values for our free parameters across our metallicity range. For the droplet radii, we chose a range from 0.01 to 100\,$\mu$m, equally spaced in log space. These radii represent the median values of the underlying log-normal distribution (geometric mean droplet radii), which is smaller than the effective droplet radii reported by \cite{AckermanMarley2001apj}. We set the width of the log-normal distribution to $1.5$ for all metallicities, as we empirically determined that this value best matches the data for all cloud species. We note that this choice naturally leads to a somewhat smaller droplet radius than if we fixed the width to the commonly used value of 1.8 \citep[see][]{AckermanMarley2001apj}. For the droplet volume mixing ratio, we chose a range from 10$^{-7}$ to 10$^{-3}$, also equally spaced in log space. Our cloud base was calculated at the intersection of the $T-P$ and $V-P$ curves for the corresponding metallicity. We assumed that the cloud extends to the top of the atmosphere and that the droplet volume mixing ratio is uniform across all cloud layers. For the planetary radius at the reference pressure of 1\,mbar, we explored a grid of planetary radii, from 2.1 to 2.7\,$\si{\rearth}$, which also includes the radius of 2.313\,$\si{\rearth}$ derived from our broadband light curve (see Sect.~\ref{subsec:pacmam}). Our detailed analysis showed that this planetary radius best fits the data; therefore, in the following text, we focus on the results derived from it.

In addition to assessing how models with different planetary radii fit the data, we also allowed for a vertical offset of the observed spectra when calculating the goodness of fit ($\chi^2_\nu$). This vertical offset ranged between $\pm 11$ ppm, corresponding to the $1\sigma$ uncertainty interval obtained from the broadband light curve fit (see Sect.~\ref{subsec:pacmam}). This few percent rescaling also accounts for uncertainty in the stellar radius. Significant instrumental offsets are not expected due to the large number of stacked transits and the consistency in transit depth across the epochs.

With each parameter set, our cloud model calculated the cloud's scattering and absorption cross-sections and optical depth. The refractive indices of MnS and MgSiO$_3$ condensate species were taken from \cite{WakefordSing2015}, while those for Fe were taken from \cite{Palik1985hocs.book}. As a test, we also computed the MnS models with the refractive indices of \citet{Kitzmann2018indices}, which yielded qualitatively the same results. The scattering and absorption efficiencies were calculated using Mie-scattering theory \citep{toon1981algorithms} by utilizing the Mie-scattering code provided by Mark Marley (private communication). This code generates pre-calculated tables of scattering and absorption efficiencies over a range of wavelengths and droplet radii, on which we then interpolated depending on our wavelength region and the cloud droplet radius (see also Fig.~\ref{fig:JB_Mieff_coeffs}). 

To generate transmission spectra including clouds, we used the {\tt PyratBay} open-source framework, which is designed for spectral synthesis and retrieval of 1D atmospheric models of planetary temperatures, species concentrations, altitude profiles, and cloud coverage \citep{Cubillos_Blecic_PyratBay2021}. This framework utilizes current knowledge of atmospheric physical and chemical processes, as well as the opacities from alkali metals, molecules, collision-induced opacities, Rayleigh scattering, and clouds. The tool can run in both forward and retrieval modes, using either self-consistent approaches or alternative parameterization methods to model atmospheric processes.

Although several molecules have spectral features in the WFC3/HST wavelength bands  \citep{MacDonaldEtal2017}, for this analysis, we only considered H$_2$O as the main and most dominant molecular source of opacity. We also assumed the presence of small aerosol particles by including both the \cite{DalgarnoWilliams1962apjRayleighH2, kurucz1970atlas} models for the H, He, and H$_2$ species and the \cite{LecavelierEtal2008aaRayleighHD189733b} Rayleigh scattering model, for which we set the enhancement factor and the scattering cross-section opacity to fixed values of 0 and -4, respectively.

\begin{figure*}[t]
    \begin{minipage}{0.5\linewidth}
        \centering
        \includegraphics[width=\linewidth]{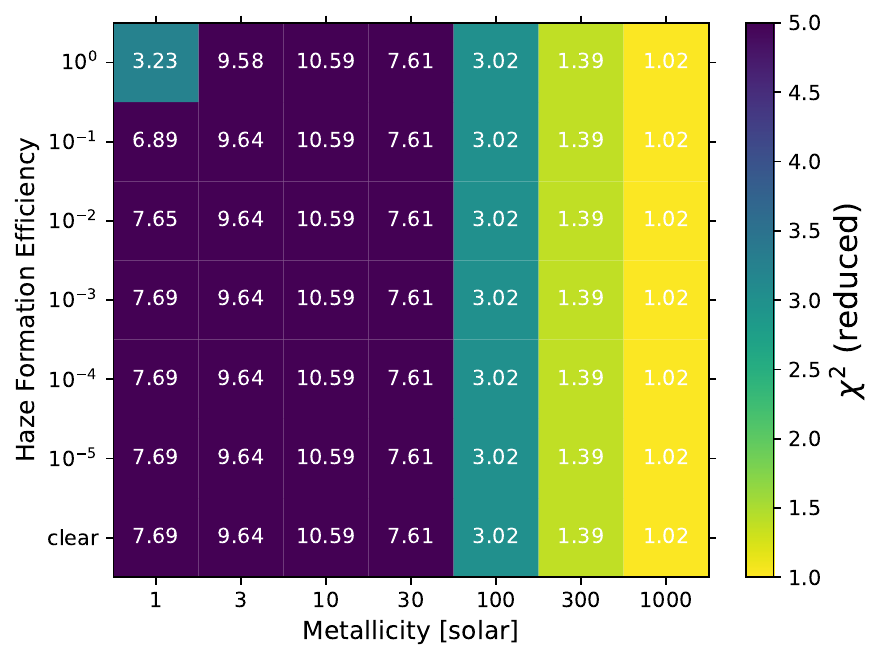}
    \end{minipage}
    \begin{minipage}{0.5\linewidth}
        \centering
        \includegraphics[width=\linewidth]{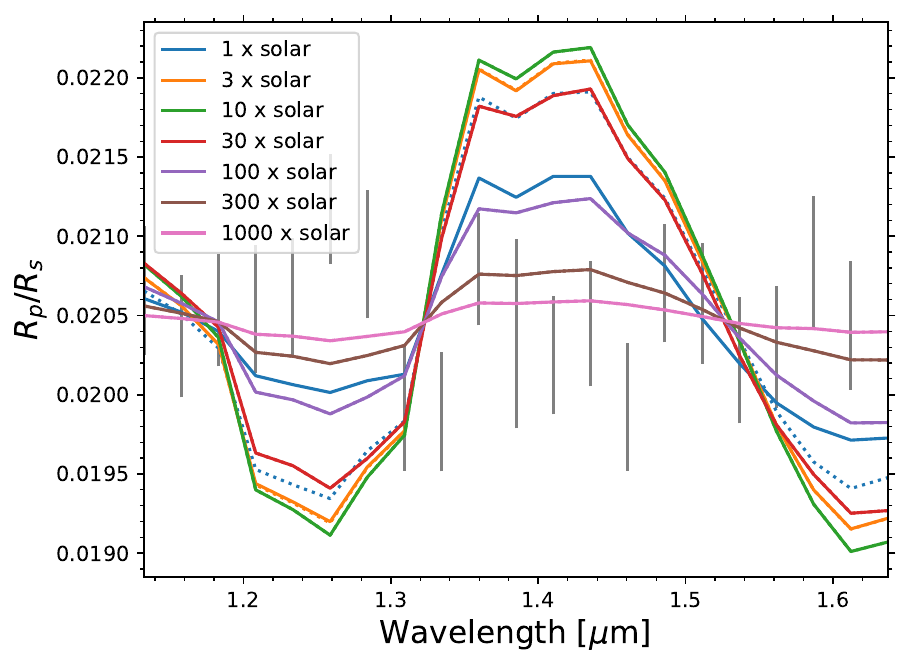}
    \end{minipage}
    \caption{Comparison of the haze models to the HST spectrum of \HD. (Left)~Goodness of fit for the parameter space (metallicity and haze formation efficiency) explored using the haze models. (Right)~Spectrum models for hazy (maximum 100\% haze formation efficiency; solid lines) and clear (dotted lines) atmospheres across the explored metallicity grid (1, 3, 10, 30, 100, 300, and 1000$\times$ solar). We note that in most metallicity cases, the solid and dotted lines overlap due to the negligible effect of haze.}
    \label{fig:haze}
\end{figure*}
Figure~\ref{fig:JB_MnS_MgSiO3_Fe} shows some of the lowest reduced chi-squared ($\chi^2_\nu$) models and the grid exploration with the goodness of fit for MnS, MgSiO$_3$, and Fe clouds. We explored the full range of our model parameters, but only the ranges that best fit the data are shown in the tables to reduce clutter. The tables list the minimum $\chi^2_\nu$ value for each set of parameters. The color bar range is set between 1 and 2 to emphasize only the models that best fit the data.

The top panels of Fig.~\ref{fig:JB_MnS_MgSiO3_Fe} explore the MnS cloud models. In the top left panel, we show several clear, no-cloud models and some representative cloud models with the lowest $\chi^2_\nu$. Only the flat model and high-metallicity models fit the data well (with reduced $\chi^{2}_{\nu}\sim 1$), while lower metallicity models require some clouds to match the data. We note that the no-cloud model at 1$\times$solar metallicity requires a high planetary radius of $R_p$ = $2.65\,\si{\rearth}$ to match the data level. This value is much larger than the value our broadband light curve analysis reports, which confirms our initial assessment that this planet either has high-altitude clouds, which increase the planetary radius, or a high metallicity atmosphere, which effectively does the same.

The middle and right panels present the $\chi^2_\nu$ values for various models across the grid. To investigate whether there are trends when exploring the grid of our model parameters, these panels depict how changing each crucial cloud parameter (while keeping the others fixed at plausible values) influences the goodness of fit. Our analysis explored all combinations of cloud parameter values across our grid, but we chose to display panels with the most $\chi^2_\nu$ values smaller than two (shown as yellow to green colors in the panels). Other parameter combinations also occasionally yielded $\chi^2_\nu <$ 2, but not as frequently as the selected examples.

The middle panel explores the goodness of fit when we keep the volume mixing ratio fixed at $q^* = 10^{-4}$ and change the droplet radius. The right panel shows models with a fixed cloud droplet radius and varying droplet volume mixing ratio, $q^*$. In general, MnS models favor higher metallicity solutions, with a slight preference for a droplet radius of $0.1\,\si{\micro\meter}$ and a droplet volume mixing ratio of $q^* = 10^{-6}$. However, the models are highly degenerate, particularly with respect to droplet radius, as different values of model parameters lead to similar values of the reduced $\chi^2_\nu$. This is independent of the fact that some models with a higher volume mixing ratio of $10^{-4}$ produce featureless spectra, while others with a lower droplet volume mixing ratio of $10^{-6}$ and smaller droplet radius of $0.1\,\si{\micro\meter}$ produce significantly less muted spectra (see orange and purple models, respectively, shown in the first panel). 

We performed a similar exploration for MgSiO$_3$ clouds (Fig.~\ref{fig:JB_MnS_MgSiO3_Fe}, middle row). MgSiO$_3$ clouds are located slightly deeper in the planetary atmosphere, around the 0.1$\,$mbar level (see Fig.~\ref{fig:JB_VP}). The lowest $\chi^2_\nu$ MgSiO$_3$ models are generally somewhat flatter than MnS clouds. Based on our grid exploration shown in the middle and right panels, the lowest $\chi^2_\nu$ models correspond to lower metallicities than MnS clouds. They also fit the data marginally better than the MnS clouds, preferring a larger volume mixing ratio, $q^* = 10^{-4}$, and a slightly smaller droplet radius, $r_g = 0.01\,\si{\micro\meter}$. We emphasize again that our $q^*$ defines the volume mixing ratio of the cloud droplets, where each droplet contains many individual cloud particles. To calculate the volume mixing ratio of the actual cloud condensates (particles), one must start with the droplet radius, account for the cloud particle radius, and consider the actual size of the droplet nucleus. For this analysis, we assumed a relatively small size for the radius of the core droplet nuclei, representing 15\% of the total droplet volume. Although the actual nucleus size is not well known, if the nuclei occupied up to 50–70\% of the droplet volume \citep{AckermanMarley2001apj}, the droplet volume mixing ratio would decrease by up to an order of magnitude, from $10^{-4}$ to $10^{-5}$ for a droplet radius of $0.01 \si{\micro\meter}$. We tested this hypothesis and observed only minor changes in the chi-squared values, with no change in the trend for any of the condensate species (as seen in Fig.~\ref{fig:JB_MnS_MgSiO3_Fe}). It is important to note that even with a nucleus size set to 15\% of the total droplet volume, for the droplet radius of $r_g = 0.01\,\si{\micro\meter}$, the droplet volume mixing ratio of $q^* = 10^{-4}$, and the MgSiO$_3$ particle radius of 2.5\,\text{\AA}, our calculated particle volume mixing ratio remains within the condensate particle volume mixing ratio reported by \citet{Morley2012TYdwarfs}, Table 2, for a solar atmospheric composition. Since most of the low $\chi^2_\nu$ models correspond to higher atmospheric metallicities, the reservoir of available MgSiO$_3$ particles would exceed that reported by \citet{Morley2012TYdwarfs} for a solar composition.
Overall, as for the MnS clouds, we reconfirm our conclusion that there is a high degeneracy between low and high-metallicity cloudy models and no-cloud high-metallicity models. 

The Fe grid model exploration led to even more degeneracies. The spectra are flatter than for MnS clouds, but $\chi^2_\nu$ values are a bit larger than for MgSiO$_3$ clouds. The Fe models with $\chi^2_\nu$ values around one spread more uniformly across all metallicities, providing no clear indication of which atmospheric metallicities might be preferred but suggesting that a slightly larger droplet radius (on the order of 1-10$\,\mu$m) and $q^* = 10^{-5}$ are needed to fit the data.

We also performed a full grid exploration assuming a planetary radius of 2.16\,$\si{\rearth}$ as given by \cite{Teske2020HD86226c}. These models, as models assuming other planetary radii, do not provide a good fit to the data. \cite{Teske2020HD86226c}'s planetary radius is derived based on only optical observations and is smaller than our broadband light curve analysis radius of 2.313\,$\si{\rearth}$. Models with the transit radius set to 2.16\,$\si{\rearth}$ have the base level of the spectrum at a much lower transit depth. To compensate for this and attempt to match the data (by lifting the spectrum and increasing the transit depth), the cloud droplet volume mixing ratio increases to high, less plausible levels (on the order of $10^{-3}$ to $10^{-2}$ for super-solar metallicities above 30$\times$ solar), strongly biasing our results toward a low-metallicity solution. 

Overall, the results show that we cannot conclusively identify which clouds condense in the atmosphere of \HD, or whether we have a clear, high metallicity atmosphere or an atmosphere with clouds. None of the considered species (MnS, MgSiO$_3$, and Fe) exhibit distinct features within the wavelength range of our observations ($1.1$ -- $1.7\,\si{\micro\meter}$), which complicates the identification of unique cloud signatures in the data. If clouds were present, their effect would be grayer and uniform across all wavelengths, which would make it impossible to identify the particular species that mutes the spectral features of \HD. 
Just marginally, our cloudy solutions match the data better than the high metallicity, no-cloud solutions. The implications of these findings are further discussed in Section \ref{sec:discussion}.

\subsection{Haze models}\label{subsec:foward_models_haze}
We also investigated the potential formation of haze in the atmosphere of \HD by simulating transmission spectra that account for the presence of haze.
The size and number density distributions of haze particles were computed using the photochemical and haze microphysical models of \citet{2018ApJ...853....7K}, following the approach of \citet{2019ApJ...877..109K}. First, we ran the photochemical model to determine the abundances of gaseous species.
In this calculation, we adopted the UV spectrum of the host star that was observed and reconstructed in Sect.~\ref{subsec:star_reduction} (see Fig.~\ref{fig:red:sed}). 
Next, we used the haze microphysical model to simulate the size and number density distributions of haze particles.
We assumed that a certain fraction of the photodissociation of precursor molecules -- CH$_4$, HCN, and C$_2$H$_2$ -- produces haze monomers.
We refer to this fraction as the haze formation efficiency.
Finally, we computed the transmission spectrum models using the derived distributions of gaseous species and haze particles.
For the optical properties of haze, we adopted the refractive index of tholin from \citet{1984Icar...60..127K}. We also tested the refractive index of soot from \citep{1998BAMS...79..831H} and found that the difference in the resulting spectra is negligible compared to our observational uncertainties. We considered the same range of metallicities as in the cloud models, namely, 1, 3, 10, 30, 100, 300, and 1000$\times$ solar, and used the same temperature-pressure profile for each case.
Other parameters, such as the eddy diffusion coefficient ($K_\mathrm{zz}$),
monomer radius ($s_1$), and internal density ($\rho_p$), were set to the same values as those used in Sect.~5.2 of \citet{Kreidberg2022HD106315b}; $K_\mathrm{zz} = 10^7$\,$\mathrm{cm}^2$\,$\mathrm{s}^{-1}$, $s_1 = 1$\,nm, and $\rho_p = 1.0$\,g\,$\mathrm{cm}^{-3}$.

The left panel of Fig.~\ref{fig:haze} shows the reduced $\chi^2$ values for the parameter space we explored, namely metallicity and haze formation efficiency.
The degrees of freedom are given by the number of data points minus the number of free parameters (metallicity and haze formation efficiency), resulting in $21 -2 = 19$.
For almost all metallicity cases, the photodissociation rates of the precursor molecules are so low that the effect of haze on the transmission spectrum is negligible. This is true even at the maximum efficiency of 100\% ($=10^0$), as shown in the right panel of Fig.~\ref{fig:haze}.
This can be explained by the relatively high atmospheric temperatures of \HD, where CO is stable instead of CH$_4$.
The low abundance of CH$_4$ also suppresses the formation of the other two haze precursors, HCN and C$_2$H$_2$, via its photodissociation, leading to negligible photodissociation rates of haze precursor molecules.
We note that the somewhat low $\chi^2$ value for the 1$\times$ solar metallicity and 100\% haze formation efficiency case is due to the relatively high abundance of CH$_4$, which results from the low temperature of the 1$\times$ solar atmosphere (see Fig.~\ref{fig:JB_VP}).
We conclude the haze formation is suppressed across the entire metallicity range. When hazes are the only considered aerosol, high-metallicity cases are favored due to their high mean molecular weight, regardless of the haze formation efficiency.

\section{Discussion}\label{sec:discussion}
The featureless spectrum of \HD continues a series of many muted sub-Neptune spectra observed in the past years~\citep[i.e.,][]{Kreidberg2014GJ1214b,Knutson2014hd97658b,Kreidberg2022HD106315b,Guo2020HD97658bmostlyflat,Wallack2024compass}. This is somewhat surprising, as observation and theory predict clearer atmospheres on sub-Neptunes with equilibrium temperatures as high as the 1310\,K of \HD (see Fig~\ref{fig:intro:SPACE}).

\begin{figure}[t]
	\centering
    \includegraphics[width=0.49\textwidth]{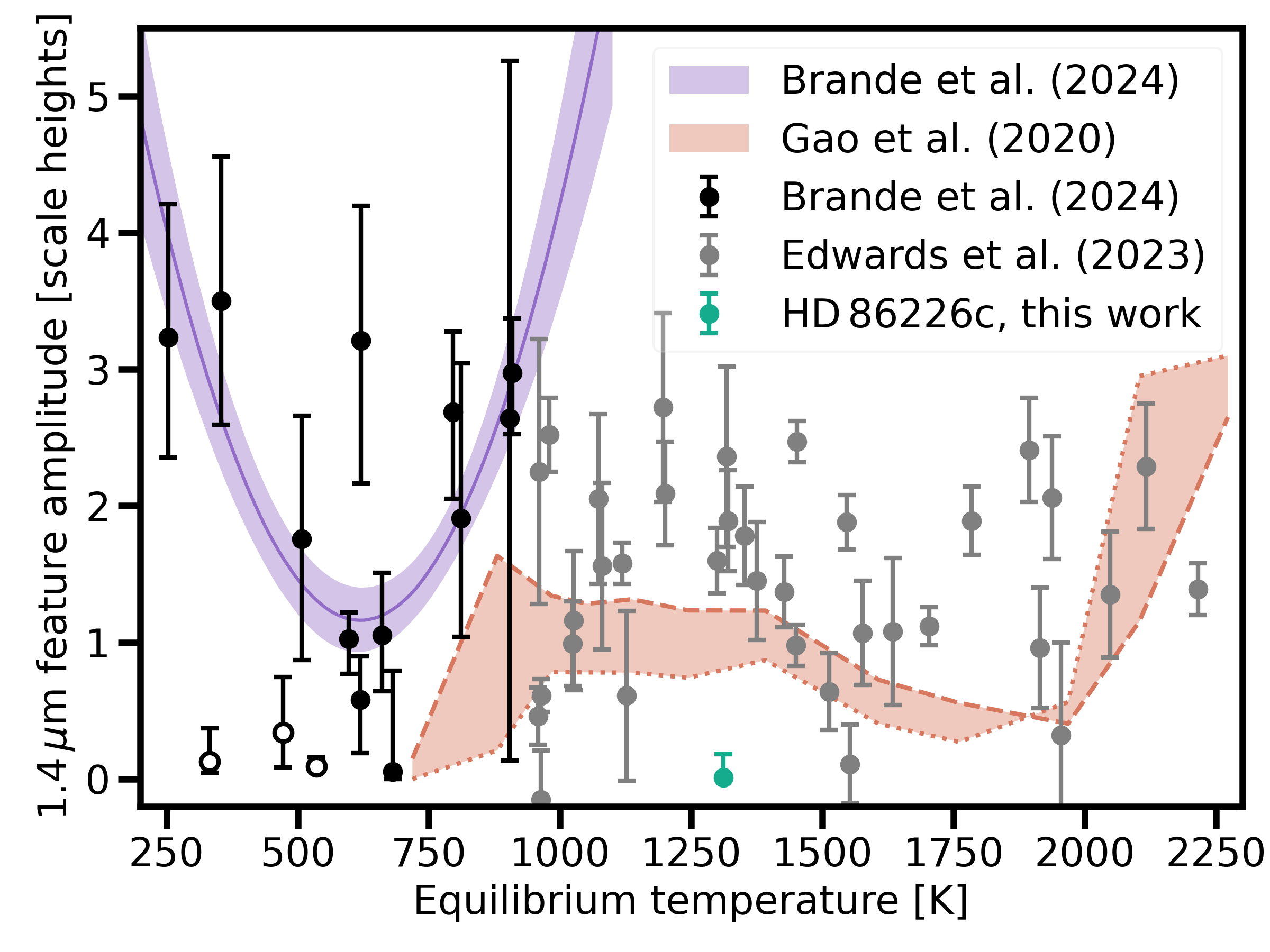}
	\caption{Trends in the $\SI{1.4}{\micro\meter}$ feature amplitude of gaseous exoplanets. The featureless spectrum of \HD (green marker) does not follow the trends seen for other planets. The purple parabolic trend was found by~\citet{Brande2024neptune_trends} and describes Neptune to sub-Neptune-sized objects (black markers). Non-filled black markers show GJ\,1214 b and the super-puffs Kepler-51 b and d, that were excluded from this trend. Gray markers show the feature amplitudes of gaseous planets calculated by~\citet{Edwards2023gasplanettrend}. Here we only include their results for planets with equilibrium temperatures above $\SI{900}{\kelvin}$, which mostly correspond to gas giants. The red region shows the feature size predictions of~\citet{Gao2020CH4} for planets with gravities of $\SI{10}{\meter\per\second\squared}$ at pressures of 1\,bar. This shaded region includes atmospheres with metallicities between $1\times$ (dotted edge) and $10\times$ (dashed edge) solar.}
	\label{fig:disc:trend}
\end{figure}

Previously characterized Neptune- to sub-Neptune sized objects show a parabolic trend in the observed $\SI{1.4}{\micro\meter}$ transmission feature size with planet equilibrium temperature~\citep[see e.g.,][]{Yu2021hazetrend,Brande2024neptune_trends}: Large features are observed at equilibrium temperatures below $\SI{400}{\kelvin}$ and above $\SI{700}{\kelvin}$. In contrast, planets with intermediate temperatures show muted spectral features. This trend is possibly shaped by hazes~\citep{Crossfield2017trend,Yu2021hazetrend} and clouds~\citep{Brande2024neptune_trends} in the planet atmospheres, which form most efficiently at intermediate temperatures around $\SI{600}{\kelvin}$. However, the trend is based on planets with equilibrium temperatures below $\SI{1000}{\kelvin}$. The featureless spectrum of \HD demonstrates that we cannot reliably extend the quadratic trend to equilibrium temperatures of $\SI{1310}{\kelvin}$ (see Fig.~\ref{fig:disc:trend}). Compared with the planets analyzed by~\citet{Brande2024neptune_trends}, the spectral feature amplitude of $0.01^{+0.17}_{-0.01}$ scale heights of \HD is among the smallest in the sample.

When giant planets are included, it is possible to extend the probed sample to equilibrium temperatures of 2500\,K. This is done by ~\citet{Fu2017gianttrend} and~\citet{Edwards2023gasplanettrend}, who find an overall increasing feature size with equilibrium temperature. They also see a local maximum in feature size at approximately $\SI{1400}{\kelvin}$ (see Fig.~\ref{fig:disc:trend}), aligned with the featureless spectrum of \HD. This discrepancy might be caused by the differences between giant planet and sub-Neptune atmospheres. While giant planets in general have approximately solar-metallicity atmospheres, recent JWST observations suggest that many sub-Neptunes have metal-rich atmospheres~\citep[e.g., GJ\,1214 b, GJ\,9827 d, TOI-270 d, and GJ\,3090 b][]{Kempton2023gj1214bPhaseC, Roy2023waterworld,Piaulet-Ghorayeb2024_GJ9827d, Benneke2024TOI270d,Ahrer2025gj3090b}. High-metallicity atmospheres have a smaller scale height, which decreases the spectral feature amplitude. In addition, their reservoir of potentially cloud-forming metals is larger. This may allow for an increased formation of clouds on sub-Neptunes, further muting their spectral features. For giant planets, \citet{Gao2020CH4} explain the local feature size maximum with a sinking of silicate clouds to higher atmospheric pressures as the equilibrium temperature of a planet decreases. In sub-Neptunes, different atmospheric metallicities and temperature structures may prevent the clouds from sinking.

\subsection{The hazy scenario}
From a theoretical perspective, the formation of organic hazes is hindered at the high temperatures of \HD~\citep[see e.g.,][]{Zahnle2009soot,morley2015models,Gao2020CH4}, eliminating one possible cause for muted spectral features. Our haze models in Sect.~\ref{subsec:foward_models_haze} agree with these findings. For \HD specifically, haze formation from the photodissociation of CH$_4$, HCN, and C$_2$H$_2$ is suppressed for all considered atmospheric compositions between $1$ and $1000\times$ solar. 

For atmospheric compositions with metallicities above $100\times$ solar, \citet{He2020haze} demonstrate that CO and CO$_2$ can also function as haze precursors. If these species also contribute to lower-metallicity atmospheres, the haze production rate could be much higher than our calculations predict. Specifically, for a $1\times$ solar atmosphere with 100\% haze formation efficiency, the haze production rate increases by four orders of magnitude from $\SI{1.0e-14}{\gram\per\centi\meter\squared\per\second}$ to $\SI{1.3e-10}{\gram\per\centi\meter\squared\per\second}$. However, it remains unclear whether CO and CO$_2$ also act as haze precursors in low-metallicity atmospheres.

Since CO and CO$_2$ are possible haze precursors, our haze models based on CH$_4$, HCN, and C$_2$H$_2$ serve as a calculation with the lower limit of the haze production rate. In this lower limit, the haze produced is not sufficient to mute the atmospheric features of \HD. If aerosols shape the muted spectrum of \HD, these are either hazes primarily formed through the photodissociation of CO and CO$_2$ or cloud species.

\subsection{The cloudy scenario}
Many cloud species may contribute to opaque high-temperature atmospheres, as shown in Fig.~\ref{fig:JB_VP}. We particularly focus on MnS, MgSiO$_3$, and Fe clouds, as the corresponding condensation curves indicate that these species could condense at the atmospheric pressures that are probed by our observations, $10^{-2}-10^{-4}\,\mathrm{bar}$. The presence of other silicate species such as Mg$_2$SiO$_4$ and SiO$_2$~\citep[see e.g.,][]{Grant2023quartz,Dyrek2024wasp107b,Inglis2024quartz} is also plausible for this planet, although these would condense at higher atmospheric pressures than MgSiO$_3$ and are likely to be less abundant~\citep[][]{lodders2006chemistry,VisscherEtal2010}. Most cloud species typically predicted on smaller Neptune-sized planets, such as Na$_2$S, KCl, and ZnS, evaporate at the temperatures of \HD. 

Our cloud models show that any of the considered species can mute the planetary spectrum of \HD. High-metallicity atmospheric solutions without clouds fit the data equally well as cloudy solutions (Section~\ref{subsec:disc:metalrich}). To reproduce the featureless spectrum of \HD with low atmospheric metallicity and MgSiO$_3$ clouds, our models require droplet volume mixing ratios of $10^{-4}$. This relatively high value results partly from our sparse grid (which is not sensitive to solutions between $10^{-4}$ and $10^{-5}$) and partly from our choice of droplet nucleus size. To verify the robustness of our results, we conducted tests using a more finely spaced grid and larger nucleus sizes. The trend observed in Fig.~\ref{fig:JB_MnS_MgSiO3_Fe} remained unchanged. Both cloudy and high-metallicity solutions stayed equally probable. 

Another point to consider is that not all of these particles may be available for cloud droplet formation. Some may participate in other chemical processes or be inhibited by rapid vertical mixing. Vertical mixing can also replenish certain species, thereby sustaining the gas-phase reservoir available for condensation. In this case, the necessary species may be supplied from deeper atmospheric layers. In contrast, larger droplets may also sink to deeper layers as a consequence of gravitational settling. Our cloud prescription applies a simplifying assumption that clouds are homogenously distributed above altitudes where the condensation curves cross the pressure-temperature profile. In the presence of large-scale zonal flows such as the ones present in hot Jupiters~\citep[see e.g.,][]{Fortney2021hotJs}, this description may be inaccurate. These flows may mix condensates below the pressure level predicted by this crossing. Furthermore, they would also preferably transport a specific size range of condensate droplets. Larger droplets would sink to lower atmospheric layers, while smaller droplets could be lofted. Depending on the position and extent of such a flow in the planet's atmosphere, a vertically homogenous cloud droplet size distribution does not reproduce this effect.

Overall, the observed featureless spectrum of \HD, along with its associated uncertainties, does not allow us to accurately constrain the cloud parameters or break the degeneracy between the cloudy and high metallicity solutions, and differentiate the actual condensate species muting the spectrum. However, these could be achieved utilizing the JWST, particularly the observations made with the MIRI instrument. Unlike Fe and MnS condensates, MgSiO$_3$ exhibits a prominent feature between 8 and 12 microns (see Fig.~\ref{fig:JB_Mieff_coeffs}), which may be observable.

\subsection{The metal-rich scenario}\label{subsec:disc:metalrich}
Recent JWST observations of the sub-Neptunes suggest that metal-rich atmospheres are common on sub-Neptunes~\citep[e.g., GJ\,1214 b, GJ\,9827 d, TOI-270 d, and GJ\,3090 b][]{Kempton2023gj1214bPhaseC, Roy2023waterworld,Piaulet-Ghorayeb2024_GJ9827d,Benneke2024TOI270d, Ahrer2025gj3090b}. The featureless spectrum of \HD can also be explained by a high mean molecular weight atmosphere alone, without the need for aerosols.

As derived in Sect.~\ref{subsec:pRT}, a cloudless atmosphere could reproduce the spectrum with a metallicity of [M/H]=$2.9^{+0.1}_{-0.2}$. This results in a $3\,\sigma$ confidence lower limit on the atmospheric metallicity of \HD of [M/H]=$2.3$, corresponding to a minimum mean molecular weight of approximately $6$. Assuming H$_2$O is the only metal in the atmosphere, this would require a water fraction of 25\% in the planet's atmosphere. Many scenarios could lead to such a high metallicity atmosphere. The atmosphere could have been enriched in volatiles by an outgassing from the planet's interior~\citep{Piette2023lavaworlds}. In this scenario, the dayside of \HD is assumed to be hot enough to evaporate the surface of the planet. In contrast, \HD could also have formed volatile-rich~\citep[see e.g.,][]{Izidoro2021pebblemigration,Bitsch2021watercontent,Burn2024valley}. Such a volatile-rich atmosphere would then be mostly stable against mass loss~\citep[][]{Lopez2017_born_dry}, explaining how the atmosphere of \HD survived until today. 

A high mean molecular weight atmosphere of \HD in combination with the presence of the outer giant companion HD\,86226\,b~\citep[][]{Teske2020HD86226c} would imply interesting scenarios for its formation history. The high mean molecular weight atmosphere could originate from the accretion of water vapor from an enriched inner disk~\citep[][]{Bitsch2021watercontent}, which might have formed from inward drifting and evaporating water pebbles~\citep{Schneider2021disk}. However, the outer growing giant beyond the water ice line could open a gap in the disk to block the inward-flowing pebbles. This would reduce the vapor enrichment in the inner regions of the disk and thus reduce the mean molecular weight of a planet forming in it. However, a high molecular weight atmosphere of \HD would indicate two possible scenarios: The giant planet could have formed interior to the water ice line of the system, where the water ice pebbles have already evaporated. Alternatively, \HD could have formed exterior to the ice line and migrated inward, before the outer giant planet could block the inward flow of pebbles~\citep[e.g.,][]{Bitsch2021watercontent}. This system is the second system after \mbox{HAT-P-11}, where such a formation analysis is possible~\citep[see][]{Chatziastros2024HAT-P-11}. Knowing the composition of the inner planets can therefore help us learn about the different formation scenarios of outer giant planets, a puzzle also present in the Solar System~\citep[e.g.,][]{Kruijer2017jupiter}.

\subsection{Possible interior structures of \HD}
Previous works reported a planetary radius of $R = 2.16 \pm 0.08\,\si{\rearth}$ for HD 86226 c \citep{Teske2020HD86226c}. Our transmission spectrum updates the radius to $R=2.313\pm0.051\,\si{\rearth}$, implying a lower bulk density. Given this density estimate and the cloud-metallicity degeneracy, \HD could have either an H/He-dominated envelope with high-altitude condensates or a metal-rich atmosphere. In the first scenario, H/He would constitute no more than 1\% in mass. H/He could form an atmospheric layer that is distinct from a rocky core \citep[i.e., a gas dwarf interior structure; see Figure 1 in][]{Lopez_Fortney14}. However, this scenario is unlikely, as planetary cores and hydrogen envelopes are chemically reactive, producing H$_2$O, CH$_4$, and other gaseous species that mix within the envelope and core \citep{schlichting22}. For high-metallicity envelopes ([M/H] $>$ 3), \HD could accommodate an envelope mass fraction of 20-50\% \citep[see Figure 5 in][]{Aguichine21}. 

The combination of mass, radius, and atmospheric metallicity alone presents a degenerate problem in constraining the size and the composition of the deep core \citep{Kramm11,Otegi20}. Therefore, the core composition could range from pure rock to a mixture of rock and iron with dissolved volatiles \citep{Vazan22,Luo24}. Determining the atmospheric abundances of H$_{2}$O, CO$_{2}$, and CH$_{4}$ with future JWST transmission spectra may help constrain the composition of the deep interior, particularly the water-to-hydrogen ratio \citep{Yang24}. However, further modeling is required to confirm whether this is also applicable to hot ($T_\U{eq} > $1000 K) sub-Neptunes similar to \HD.

\subsection{How to solve the cloud-metallicity degeneracy}
The only atmospheric scenario that can be unequivocally ruled out with the HST spectrum of \HD is a cloud-free, solar metallicity atmosphere. Based on our atmospheric retrievals in Sect.~\ref{subsec:pRT}, we can exclude this model with a significance of $6.5\,\sigma$. Nevertheless, we cannot conclusively identify if clouds condense in the atmosphere of \HD, or whether it has a clear, high-metallicity atmosphere. The atmosphere of \HD could also combine high-altitude clouds and an enhanced metallicity.

\begin{figure}[t]
	\centering
	\includegraphics[width=0.49\textwidth]{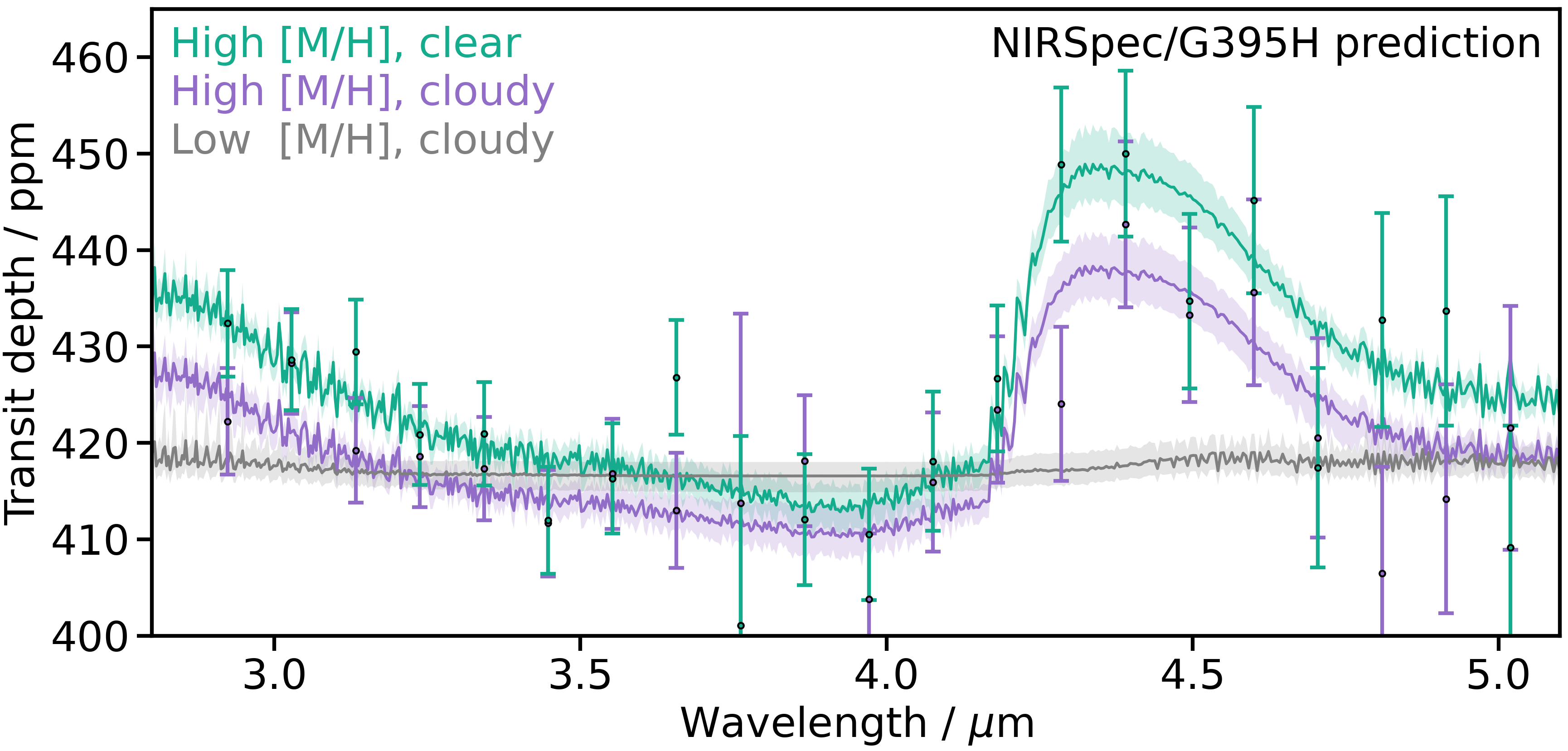}
	\caption{Simulated JWST spectra for different atmospheric compositions of \HD. The colored data points show JWST NIRSpec/G395H mock data of three transits, simulated with PandExo~\citep[][]{Batalha2017pandexo}. Colored contours show $1\,\sigma$ percentiles drawn from the posterior distributions of the pRT retrievals on the simulated data. Simulated spectra are based on three edge-case atmospheric scenarios that are feasible based on the pRT fits to the HST data in Sect.~\ref{subsec:pRT}: high-metallicity ($[\mathrm{M}/\mathrm{H}]>2.5$) solutions with a cloud-deck below -$1$\,bar (green), high-metallicity solutions with a cloud-deck above $-3$\,bar (purple), and low-metallicity ($[\mathrm{M}/\mathrm{H}]<1$) solutions with a cloud-deck above $-3$\,bar (gray, data points removed to avoid clutter).}
	\label{fig:disc:jwst}
\end{figure}

This degeneracy could be broken by observing \HD with the NIRSpec/G395H instrument of JWST, as CO$_2$ has a prominent feature above 3\,$\mu $m that is highly sensitive to the atmospheric metallicity. Figure~\ref{fig:disc:jwst} displays the predicted spectra for different possible atmospheric edge-cases on \HD, as derived from the HST data: high-metallicity ($[\mathrm{M}/\mathrm{H}]>2.5$) atmospheres with a gray cloud-deck below $-1$\,bar or above $-3$\,bar, and a low-metallicity ($[\mathrm{M}/\mathrm{H}]<1$) atmosphere with a cloud-deck above $-3$\,bar. For each of these scenarios, we used pRT to generate spectra from all retrieval posteriors (see Sect.~\ref{subsec:pRT}) that match the respective description. For these spectra, we included the opacities of H$_2$O, CO, and CO$_2$, as these are the dominant species predicted for planets similar to \HD by~\cite{Moses2013composition}. We used the median of these spectra to obtain a model spectrum for each scenario, respectively. For each of these three model spectra, we used \texttt{pandexo}~\citep[][]{Batalha2017pandexo} to simulate NIRSpec/G395H data of three transit observations. On the simulated data, we additionally added a $1\sigma$ random noise. 

To verify that the three scenarios can be distinguished, we retrieved the simulated spectrum with pRT. Assuming scaled solar abundances, high-metallicity solutions show a detectable CO$_2$ spectral feature. This species does not contribute to the observable opacity of a low-metallicity atmosphere, which allows us to differentiate between the scenarios. While we only included CO, CO$_2$, and H$_2$O in our predictions, OCS could be another opacity contributor that traces metal-rich atmospheres~\cite{Moses2013composition}. Depending on the exact composition of the atmosphere, the amplitude of the $\SI{4.8}{\micro\meter}$ OCS feature could be similar to that of CO$_2$. The cloud opacity usually diminishes in the wavelength region between 3 to 5$\,\si{\micro\meter}$~\citep[e.g.,][]{Benneke2024TOI270d,Beatty2024So2_GJ3470b,Schlawin2024}, facilitating a measurement of the atmospheric metallicity. 

Observations with JWST/MIRI may enable a better distinction of the cloud species that mute the spectra of \HD, by probing the clouds directly. If silicate clouds are present in \HD, it may be possible to observe the silicate features around $8-10\,\si{\micro\meter}$~\citep[see Fig.~\ref{fig:JB_Mieff_coeffs} and][]{Marley2015clouds}. However, directly detecting silicate features in the possibly metal-enriched atmosphere of \HD may require much observing time, as previously simulated by~\citet{Piette2023lavaworlds}. At shorter wavelengths, Rayleigh scattering on small cloud particles might produce a spectral slope. High sensitivity observations at visual wavelengths could therefore provide a further probe of the cloud coverage on \HD. Finally, another possibility for detecting and characterizing clouds on the planet is to measure the occultation depth, or phase curve, of the planet~\citep[see e.g.,][]{Kempton2023gj1214bPhaseC,Coulombe2025LTT9779bPhaseC}. In the presence of silicate clouds, the planet is expected to have a high albedo, reflecting a large fraction of the incoming radiation~\citep[e.g.,][]{Garcia2015phaseC}.

\section{Summary}\label{sec:summary}
To characterize the atmosphere of the hot sub-Neptune HD\,86226\,c, we observed nine transits of the planet with HST/WFC3 over the wavelength range 1.1 -- 1.7 $\mu$m. We also measured the UV spectrum of the host star with HST/STIS and reconstructed the panchromatic stellar spectrum with scaling relationships and stellar models. In addition, we performed photometric monitoring with the $0.6$\,m telescope at the Van Vleck Observatory.

For the HST/WFC3 data reduction, we used the open-source \texttt{PACMAN} and \texttt{Eureka!} packages. In contrast to typical WFC3 spatial scanning observations, the high scan rate for this bright star resulted in shifts in the position of the spectral trace on the detector. We observed shifts of up to two pixels, resulting in a flux measurement that was correlated with the position of the trace. To account for this effect, we implemented a function in \texttt{PACMAN} that decorrelates flux and trace position for both forward and reverse scan directions. For the data reduction with \texttt{Eureka!}, we modified the pipeline to allow for the simultaneous fitting of multiple HST transits. We find that \texttt{PACMAN} and \texttt{Eureka!} result in qualitatively similar spectra, but we note that \texttt{Eureka!} does not correct the trace position instabilities, resulting in a higher level of correlated noise in the light curves.

The optical photometric monitoring revealed HD\,86226 as a quiet host star with a standard deviation in nightly variability of $31 \,\mathrm{mmag}$ in the V band. Our nine transit observations do not show evidence of stellar flares or spot crossings of the planet.

The transmission spectrum of HD\,86266\,c is featureless and consistent within $0.4\,\sigma$ with a constant transit depth of $418\pm14$\,ppm. This corresponds to an amplitude of just $0.01$ scale heights for an H/He-dominated atmosphere. Based on an atmospheric retrieval analysis with petitRADTRANS, we ruled out a cloud-free solar metallicity atmosphere with a confidence of $6.5\,\sigma$. However, the existing data cannot unambiguously distinguish whether aerosols or a high-metallicity atmosphere are responsible for the muted spectral features. In the absence of clouds, the observed spectrum could be produced by a metal-enriched atmosphere with $[\mathrm{M}/\mathrm{H}]>2.3$ ($3\,\sigma$ confidence lower limit). In addition to the atmospheric retrieval analysis, we also compared the spectrum to predictions from forward models of clouds and hazes. In contrast to most sub-Neptunes, we find that the high temperature of HD\,86226\,c prohibits the formation of organic hazes with CH$_4$, HCN, and C$_2$H$_2$ as haze precursors. On the other hand, it allows for the formation of refractory cloud species such as MnS, MgSiO$_3$, and Fe clouds, which are similar to those expected for hot Jupiters. All of these cloud species could plausibly match the observations.

The featureless spectrum of the hot and small sub-Neptune \HD is surprising in the context of previously observed trends for sub-Neptune atmospheres, namely that hotter planets should have larger amplitude spectral features \citep[e.g.,][]{Brande2024neptune_trends}. Follow-up observations with JWST may clarify the physical processes and atmospheric chemistry that shape the observed spectrum. Determining this planet's exact atmospheric composition will show if it hosts a cloud species atypical for sub-Neptunes or if it was formed with a metal-enriched atmosphere.

\section{Data availability}
The HST data used in this paper are associated with the program GO 17192 (P.I., L.~Kreidberg) and are publicly available via
the Mikulski Archive for Space Telescope at \url{https://mast.stsci.edu}. Additional data products derived in this work are available via Zenodo at \url{https://zenodo.org/records/16035649}.

\begin{acknowledgements}
    This research is based on observations made with the NASA/ESA Hubble Space Telescope obtained from the Space Telescope Science Institute, which is operated by the Association of Universities for Research in Astronomy, Inc., under NASA contract NAS 5–26555. These observations are associated with program 17192. We wish to thank Peter Gao for sharing the aerosol model spectra with us. Furthermore, we thank the anonymous referee for the detailed report that led to the investigation of many important details and a better overall clarity of the text. Q.C.T.~and S.R.~wish to thank Roy Kilgard for his assistance in the operation of the Wesleyan 24-inch telescope and the data storage system. K.A.K.~gratefully acknowledges support from the DLR via project P.S.ASTR1508. Y.K.~acknowledges support from JSPS KAKENHI Grant Numbers 21K13984, 22H05150, and 23H01224.
    J.S.J.~greatfully acknowledges support by FONDECYT grant 1240738 and from the ANID BASAL project FB210003. P.E.C. acknowledges ﬁnancial support by the Austrian Science Fund (FWF) Erwin Schroedinger Fellowship, program J4595-N. Portions of this research were carried out on the High Performance Computing resources at New York University Abu Dhabi. T.D. acknowledges support from the McDonnell Center for the Space Sciences at Washington University in St. Louis.
\end{acknowledgements}
\bibliographystyle{aa} 
\bibliography{aa54916-25} 
\begin{appendix}
\section{\texttt{PACMAN} light curve fits }\label{app:allan_pacman}
Table~\ref{app:tab:transit_depth_pacman} lists the transit depths derived in Sect.~\ref{sec:LC} from the light curve fits with \texttt{pacman}. Figures showing the behavior of the RMS when the individual HST exposures are time-sorted and binned~\citep[see e.g.,][]{Cubillos2017beta} are available via \href{https://zenodo.org/records/16035649}{Zenodo}. The broadband light curve shows residual red noise, which could be removed by fitting a second-order baseline in time to the flux of each visit. As argued in Sect.~\ref{subsec:pacmam}, this was not done to avoid overfitting the individual wavelength bins of the spectroscopic light curves and to have a common systematics model for broadband and spectroscopic light curves. To estimate the uncertainty of the broadband transit depth, we applied an uncertainty scaling with a constant $\beta$-factor of 2, as described by~\citet{Pont2006beta} and~\citet{Cubillos2017beta}. This factor was derived from the fraction between fit RMS and expected white-noise RMS~\citep[see e.g.,][]{Cubillos2017beta} at a bin size of 25 exposures. This bin size corresponds to the duration of one HST orbit, the longest duration of continuous observations in our dataset. 

For the spectroscopic bins, the increased RMS makes the deviation from a linear flux baseline undetectable, which is why the RMS behaves similar to white noise in most wavelength bins. Minor deviations from this are seen in the bin at $\SI{1.31}{\micro\meter}$, as there is an artifact on the WFC3 detector. Furthermore, bins at the long-wavelength edge show slight signs of red noise. Since this contribution is minor, we did not account for it further. 

\begin{table}[htb]
    \centering
        \caption{Transit depths of HD\,86226\,c derived with \texttt{PACMAN}.}
        \label{app:tab:transit_depth_pacman}
    \begin{tabular}{ccc}
        \hline
        Wavelength & Transit Depth & Uncertainty \\ 
        {\(\mu\text{m}\)} & ppm & ppm \\
        \hline
        Broadband & 410 & 11 \\
        1.133 & 427 & 17 \\ 
        1.158 & 415 & 16 \\ 
        1.183 & 422 & 15 \\ 
        1.208 & 422 & 17 \\ 
        1.234 & 425 & 15 \\ 
        1.259 & 448 & 15 \\ 
        1.284 & 436 & 17 \\ 
        1.309 & 397 & 16 \\ 
        1.335 & 396 & 15 \\ 
        1.360 & 433 & 15 \\ 
        1.385 & 416 & 24 \\ 
        1.410 & 410 & 15 \\ 
        1.435 & 418 & 16 \\ 
        1.461 & 397 & 16 \\ 
        1.486 & 429 & 15 \\ 
        1.511 & 424 & 16 \\ 
        1.536 & 409 & 16 \\ 
        1.562 & 412 & 16 \\ 
        1.587 & 434 & 17 \\ 
        1.612 & 418 & 17 \\ 
        1.637 & 395 & 18 \\ 
        \hline
        \hline
    \end{tabular}
    \tablefoot{The uncertainties of the broadband light curve correspond to the $1\,\sigma$ percentiles from the \texttt{PACMAN} output multiplied by the $\beta$-factor of 2.}
\end{table}

\section{\texttt{Eureka!} light curve fits }\label{app:allan_eureka}
Table~\ref{app:tab:transit_depth_eureka} lists the transit depths derived in Sect.~\ref{sec:LC} from the light curve fits with \texttt{Eureka!}. Plots of the corresponding light curves are available via \href{https://zenodo.org/records/16035649}{Zenodo}. With \texttt{Eureka!}, it was not possible to accurately describe the exact HST ramp behavior. This issue was accounted for in the fits to the spectroscopic bins by using the \texttt{divide-white} technique~\citep[e.g.,][]{Kreidberg2014GJ1214b}. 

The spectroscopic light curve fits with \texttt{Eureka!} show red noise in most wavelength bins and the broadband light curve (the corresponding figures are available at \href{https://zenodo.org/records/16035649}{Zenodo}). As explained further in Sects.~\ref{subsec:eureka} and~\ref{subsec:PACMANandEureka}, this red noise mainly emerges from the use of the first orbit of every visit and the poor stability of the trace position on the detector during the observations. We accounted for this remaining red noise according to the method described by~\citet{Pont2006beta} and~\citet{Cubillos2017beta}: The uncertainty on the transit depth of each wavelength bin was inflated by a bin-specific $\beta$-factor. This factor was derived from the fraction between fit RMS and expected white noise RMS~\citep[see e.g.,][]{Cubillos2017beta} at a bin size of 25 exposures. This bin size corresponds to the duration of one HST orbit, the longest duration of continuous observations in our dataset. 

\begin{table}[htb]
    \centering
        \caption{Transit depths of HD\,86226\,c derived with \texttt{Eureka!}.}
        \label{app:tab:transit_depth_eureka}
    \begin{tabular}{cccc}
        \hline
        Wavelength & Transit Depth & Uncertainty & $\beta$\\ 
        {\(\mu\text{m}\)} & ppm & ppm & \\
        \hline
        Broadband & 460 & 17 & 2.3\\
        1.13 & 636 & 66 & 2.1 \\ 
        1.15 & 464 & 42 & 2.5 \\ 
        1.17 & 404 & 49 & 2.6 \\ 
        1.19 & 414 & 46 & 2.8 \\ 
        1.21 & 526 & 57 & 2.6 \\ 
        1.23 & 378 & 44 & 2.8 \\ 
        1.25 & 465 & 29 & 2.3 \\ 
        1.27 & 412 & 22 & 2.0 \\ 
        1.29 & 518 & 26 & 1.9 \\ 
        1.31 & 433 & 23 & 1.5 \\ 
        1.33 & 422 & 25 & 1.7 \\ 
        1.35 & 439 & 17 & 1.5 \\ 
        1.37 & 467 & 16 & 1.4 \\ 
        1.39 & 501 & 22 & 1.5 \\ 
        1.41 & 410 & 22 & 1.5 \\ 
        1.43 & 440 & 27 & 1.9 \\ 
        1.45 & 465 & 18 & 1.4 \\ 
        1.47 & 483 & 22 & 1.4 \\ 
        1.49 & 459 & 24 & 1.5 \\ 
        1.51 & 427 & 18 & 1.4 \\ 
        1.53 & 437 & 22 & 1.6 \\ 
        1.55 & 477 & 16 & 1.3 \\ 
        1.57 & 371 & 23 & 1.8 \\ 
        1.59 & 467 & 14 & 1.3 \\ 
        1.61 & 416 & 24 & 1.5 \\ 
        1.63 & 438 & 32 & 1.9 \\ 
        1.65 & 413 & 65 & 2.8 \\ 
        \hline
        \hline
    \end{tabular}
    \tablefoot{This uncertainties in this table correspond to the $1\,\sigma$ percentiles from the \texttt{Eureka!} output multiplied by the $\beta$-factor given in the last column. We did not include the 1.13\,$\mu$m bin in our analysis.}
\end{table}

\end{appendix}
\end{document}